\documentclass[sigconf]{acmart}

\usepackage[english]{babel}
\usepackage[utf8x]{inputenc}
\usepackage{amsmath, bm}
\usepackage{amsthm}
\usepackage{amsfonts}
\usepackage[colorinlistoftodos]{todonotes}
\usepackage{booktabs}
\usepackage{multirow}
\usepackage{enumitem}
\usepackage{mleftright}
\usepackage{diagbox}
\usepackage{booktabs,subcaption,amsfonts,dcolumn}
\usepackage[export]{adjustbox}
\usepackage{changepage}

\newcommand{\vecx}{{\bf x}}
\newcommand{\vecy}{{\bf y}}
\newcommand{\vecw}{{\bf w}}
\newcommand{\vecb}{{\bf b}}
\newcommand{\vech}{{\bf h}}
\newcommand{\vecv}{{\bf v}}
\newcommand{\veci}{{\bf i}}
\newcommand{\vecj}{{\bf j}}
\newcommand{\vece}{{\bf e}}
\newcommand{\vecalpha}{{\bm \alpha}}
\newcommand{\vecbeta}{{\bm \beta}}

\newcommand{\Var}{\mathrm{Var}}

\newcommand{\veczero}{{\bf 0}}

\addto\extrasenglish{%
}
\AtBeginDocument{%
  \providecommand\BibTeX{{%
    \normalfont B\kern-0.5em{\scshape i\kern-0.25em b}\kern-0.8em\TeX}}}

\renewcommand\footnotetextcopyrightpermission[1]{} 
\begin{document}

\title{DCN V2: Improved Deep \& Cross Network and Practical Lessons for Web-scale Learning to Rank Systems}

\author{Ruoxi Wang, Rakesh Shivanna, Derek Z. Cheng, Sagar Jain, Dong Lin, Lichan Hong, Ed H. Chi \\ Google Inc. \\
\texttt{\{ruoxi, rakeshshivanna, zcheng, sagarj, dongl, lichan, edchi\}@google.com}
}



\begin{abstract}

Learning effective feature crosses is the key behind building recommender systems. However, the sparse and large feature space requires exhaustive search to identify effective crosses. Deep \& Cross Network (DCN) was proposed to automatically and efficiently learn bounded-degree predictive feature interactions. Unfortunately, in models that serve web-scale traffic with billions of training examples, DCN showed limited expressiveness in its cross network at learning more predictive feature interactions. Despite significant research progress made, many deep learning models in production still rely on traditional feed-forward neural networks to learn feature crosses inefficiently.

In light of the pros/cons of DCN and existing feature interaction learning approaches, we 
propose an improved framework {DCN-V2} to make DCN more practical in large-scale industrial settings. In a comprehensive experimental study with extensive hyper-parameter search and model tuning, we observed that {DCN-V2} approaches outperform all the state-of-the-art algorithms on popular benchmark datasets. The improved {DCN-V2} is more expressive yet remains cost efficient at feature interaction learning, especially when coupled with a mixture of low-rank architecture. {DCN-V2} is simple, can be easily adopted as building blocks, and has delivered significant offline accuracy and online business metrics gains across many web-scale learning to rank systems at Google.

\end{abstract}


\pagestyle{plain} 


\maketitle

\section{Introduction}

Learning to rank (LTR) \cite{liu2011learning, cao2007learning} has remained to be one of the most important problems in modern-day machine learning and deep learning. It has a wide range of applications in search, recommendation systems \cite{resnick1997recommender, herlocker2004evaluating, schafer1999recommender}, and computational advertising \cite{broder2008computational, bottou2013counterfactual}. Among the crucial components of LTR models, learning effective feature crosses continues to attract lots of attention from both academia \cite{qu2016product, lian2018xdeepfm, song2019autoint} and industry \cite{wang2017deep, cheng2016wide, guo2017deepfm, beutel2018latent, naumov2019deep}.

Effective feature crosses are crucial to the success of many models. They provide additional interaction information beyond individual features. For example, the combination of ``{\ttfamily{country}}'' and ``{\ttfamily{language}}'' is more informative than either one of them.
In the era of linear models, ML practitioners rely on manually identifying such feature crosses \cite{seide2011feature} to increase model's expressiveness.
Unfortunately, this involves a combinatorial search space, which is large and sparse in web-scale applications where the data is mostly categorical.
Searching in such setting is exhaustive, often requires domain expertise, and makes the model harder to generalize.

Later on, embedding techniques have been widely adopted to project features from high-dimensional sparse vectors to much lower-dimensional dense vectors. Factorization Machines (FMs) \cite{rendle2010factorization, rendle:tist2012} leverage the embedding techniques and construct pairwise feature interactions via the inner-product of two latent vectors. Compared to those traditional feature crosses in linear models, FM brings more generalization capabilities.

In the last decade, with more computing firepower and huge scale of data, LTR models in industry have gradually migrated from linear models and FM-based models to deep neural networks (DNN). This has significantly improved model performance for search and recommendation systems across the board \cite{cheng2016wide, wang2017deep, guo2017deepfm}. People generally consider DNNs as universal function approximators, that could potentially learn all kinds of feature interactions \cite{mhaskar1996neural, valiant2014learning, NIPS2016_6556}. However, recent studies \cite{beutel2018latent, wang2017deep} found that DNNs are inefficient to even approximately model 2nd or 3rd-order feature crosses.

To capture effective feature crosses more accurately, a common remedy is to further increase model capacity through wider or deeper networks. This naturally crafts a double edged sword that we are improving model performance while making models much slower to serve. In many production settings, these models are handling extremely high QPS, thus have very strict latency requirements for real-time inference. Possibly, the serving systems are already pushed to a stretch that cannot afford even larger models. Furthermore, deeper models often introduce trainability issues, making models harder to train.

This has shed light on critical needs to design a model that can efficiently and effectively learn predictive feature interactions, especially in a resource-constraint environment that handles real-time traffic from billions of users. Many recent works \cite{wang2017deep, cheng2016wide, guo2017deepfm, beutel2018latent, qu2016product, lian2018xdeepfm, song2019autoint, naumov2019deep} tried to tackle this challenge. The common theme is to leverage those \emph{implicit} high-order crosses learned from DNNs, with \emph{explicit} and bounded-degree feature crosses which have been found to be effective in linear models. \emph{Implicit} cross means the interaction is learned through an end-to-end function without any explicit formula modeling such cross. \emph{Explicit} cross, on the other hand, is modeled by an explicit formula with controllable interaction order. We defer a detailed discussion of these models in \autoref{sec:related_work}.

Among these, Deep \& Cross Network (DCN) \cite{wang2017deep} is effective and elegant, however, productionizing DCN in large-scale industry systems faces many challenges. The expressiveness of its cross network is limited. The polynomial class reproduced by the cross network is only characterized by $O(\text{input size})$ parameters, largely limiting its flexibility in modeling random cross patterns. 
Moreover, the allocated capacity between the cross network and DNN is unbalanced. This gap significantly increases when applying DCN to large-scale production data. An overwhelming portion of the parameters will be used to learn implicit crosses in the DNN.

In this paper, we propose a new model \emph{{DCN-V2}} that improves the original DCN model. We have already successfully deployed {DCN-V2} in quite a few learning to rank systems across Google with significant gains in both offline model accuracy and online business metrics. {DCN-V2} first learns explicit feature interactions of the inputs (typically the embedding layer) through cross layers, and then combines with a deep network to learn complementary implicit interactions. The core of {DCN-V2} is the cross layers, which inherit the simple structure of the cross network from DCN, however significantly more expressive at learning explicit and bounded-degree cross features. The paper studies datasets with clicks as positive labels, however {DCN-V2} is label agnostic and can be applied to any learning to rank systems.
The main contributions of the paper are five-fold:

\begin{itemize}[leftmargin=1em]
    \item We propose a novel model---{DCN-V2}---to learn effective explicit and implicit feature crosses. Compared to existing methods, our model is more expressive yet remains efficient and simple.
    \item Observing the low-rank nature of the learned matrix in {DCN-V2}, we propose to leverage low-rank techniques to approximate feature crosses in a subspace for better performance and latency trade-offs. In addition, we propose a technique based on the Mixture-of-Expert architecture \cite{shazeer2017outrageously, jacobs1991adaptive} to further decompose the matrix into multiple smaller sub-spaces. These sub-spaces are then aggregated through a gating mechanism.
    \item We conduct and provide an extensive study using synthetic datasets, which demonstrates the inefficiency of traditional ReLU-based neural nets to learn high-order feature crosses.
    \item Through comprehensive experimental analysis, we demonstrate that our proposed {DCN-V2} models significantly outperform SOTA algorithms on Criteo and MovieLen-1M benchmark datasets.
    \item We provide a case study and share lessons in productionizing {DCN-V2} in a large-scale industrial ranking system, which delivered significant offline and online gains.
\end{itemize}

The paper is organized as follows. \autoref{sec:related_work} summarizes related work. \autoref{sec:dcn-m} describes our proposed model architecture {DCN-V2} along with its memory efficient version. \autoref{sec:dcn_analysis} analyzes {DCN-V2}. \autoref{sec:research_qs} raises a few research questions, which are answered through comprehensive experiments on both synthetic datasets in \autoref{sec:exp_synthetic} and public datasets in \autoref{sec:exp_public}. \autoref{sec:productionization} describes the process of productionizing {DCN-V2} in a web-scale recommender. 

\section{Related Work}
\label{sec:related_work}


The core idea of recent feature interaction learning work is to leverage both explicit and implicit (from DNNs) feature crosses. To model explicit crosses, most recent work introduces multiplicative operations ($x_1 \times x_2$) which is inefficient in DNN, and designs a function $f(\vecx_1, \vecx_2)$ to efficiently and explicitly model the pairwise interactions between features $\vecx_1$ and $\vecx_2$. We organize the work in terms of how they combine the explicit and implicit components.

{\bf Parallel Structure.} One line of work jointly trains two parallel networks inspired from
the wide and deep model \cite{cheng2016wide}, where the wide component takes inputs as crosses of raw features; and the deep component is a DNN model. However, selecting cross features for the wide component falls back to the feature engineering problem for linear models. Nonetheless, the wide and deep model has inspired many works to adopt this parallel architecture and improve upon the wide component. 

DeepFM \cite{guo2017deepfm} automates the feature interaction learning in the wide component by adopting a FM model. DCN \cite{wang2017deep} introduces a cross network, which learns explicit and bounded-degree feature interactions automatically and efficiently. 
xDeepFM \cite{lian2018xdeepfm} increases the expressiveness of DCN by generating multiple feature maps, each encoding all the pairwise interactions between features at current level and the input level. Besides, it also considers each feature embedding $\vecx_i$ as a unit instead of each element $x_i$ as a unit. Unfortunately, its computational cost is significantly high (10x of \#params), making it impractical for industrial-scale applications. Moreover, both DeepFM and xDeepFM require all the feature embeddings to be of equal size, yet another limitation when applying to industrial data where the vocab sizes (sizes of categorical features) vary from $O(10)$ to millions. AutoInt \cite{song2019autoint} leverages the multi-head self-attention mechanism with residual connections. InterHAt \cite{li2020interpretable} further employs Hierarchical Attentions.

{\bf Stacked Structure.} Another line of work introduces an interaction layer---which creates explicit feature crosses---in between the embedding layer and a DNN model. This interaction layer captures feature interaction at an early stage, and facilitates the learning of subsequent hidden layers. Product-based neural network (PNN) \cite{qu2016product} introduces inner (IPNN) and outer (OPNN) product layer as the pairwise interaction layers. One downside of OPNN lies in its high computational cost. 
Neural FM (NFM) \cite{he2017neural} extends FM by replacing the inner-product with a Hadamard product;
DLRM \cite{naumov2019deep} follows FM to compute the feature crosses through inner products;
These models can only create up to 2nd-order explicit crosses. AFN \cite{cheng2019adaptive} transforms features into a log space and adaptively learns arbitrary-order feature interactions. Similar to DeepFM and xDeepFM, they only accept embeddings of equal sizes.

Despite many advances made, our comprehensive experiments (\autoref{sec:exp_public}) demonstrate that DCN still remains to be a strong baseline. We attribute this to its simple structure that has facilitated the optimization. However, as discussed, its limited expressiveness has prevented it from learning more effective feature crosses in web-scale systems. In the following, we present a new architecture that inherits DCN's simple structure while increasing its expressiveness.



\section{Proposed Architecture: {DCN-V2}}
\label{sec:dcn-m}

\begin{figure}[htbp]
\centering
    \begin{subfigure}[b]{0.2\textwidth}  
    \includegraphics[width=\textwidth]{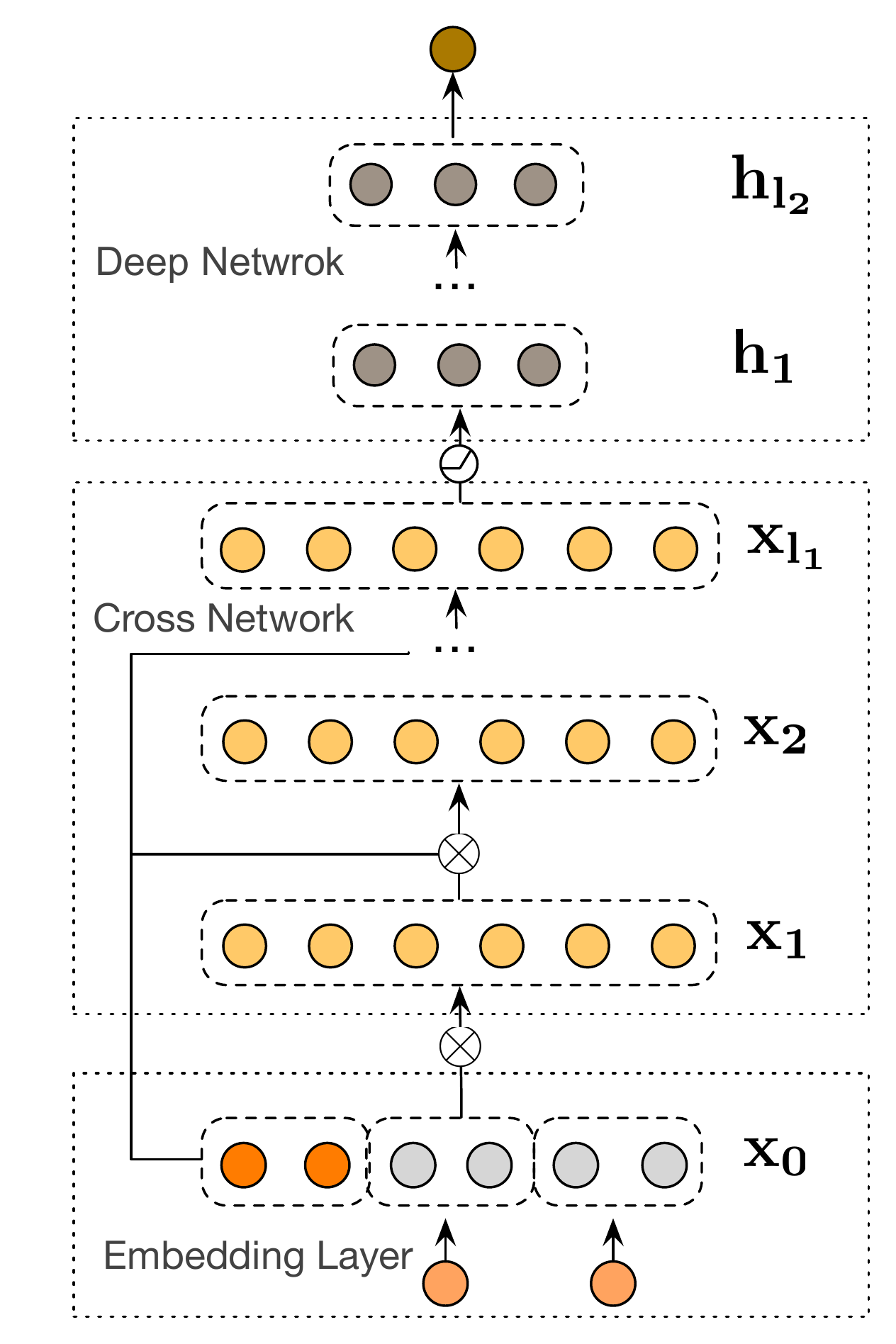}
    \caption{Stacked}
    \label{fig:dcn-stack}
    \end{subfigure}
    \hfill
    \begin{subfigure}[b]{0.27\textwidth}  
    \includegraphics[width=\textwidth]{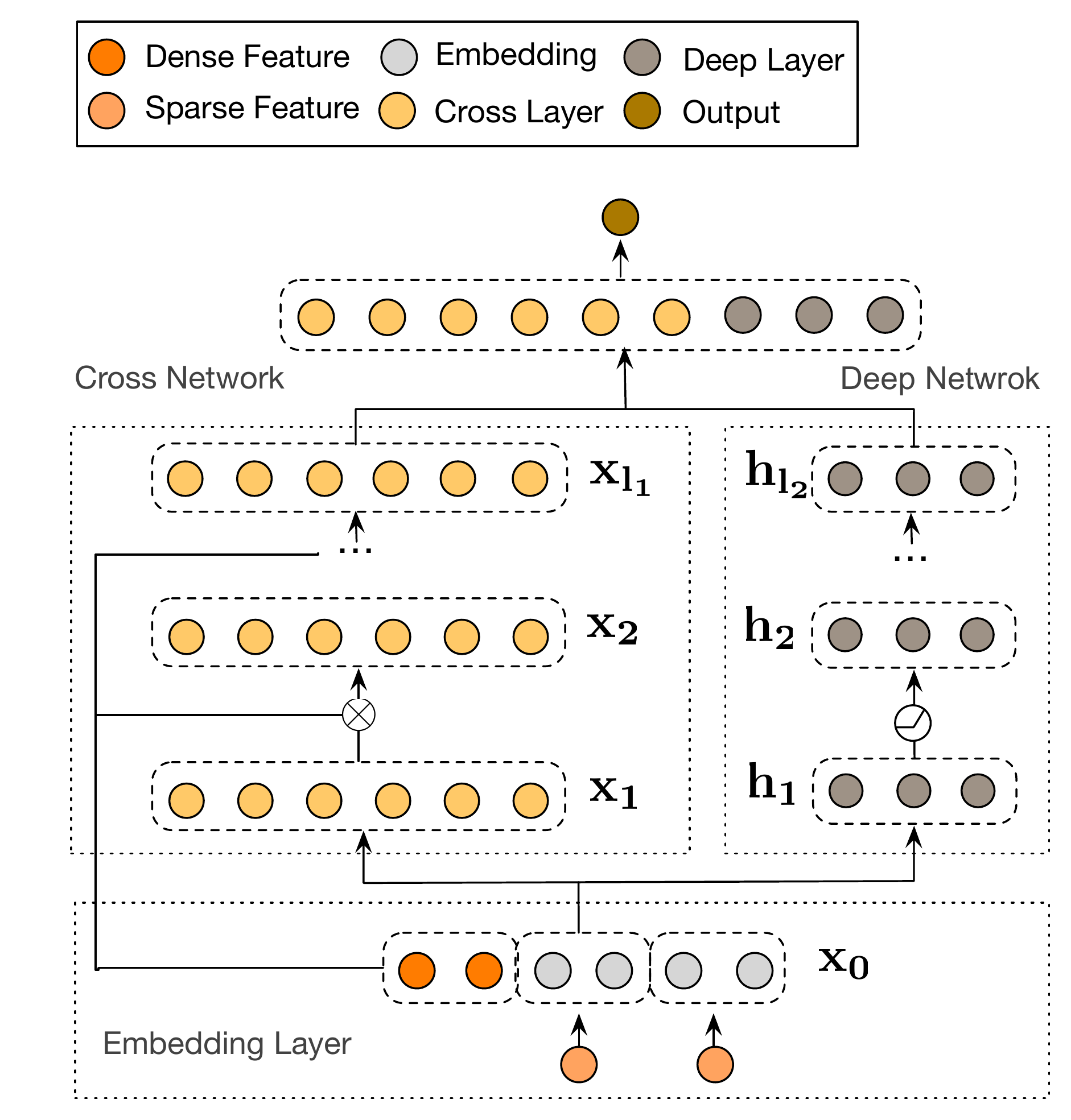}
    \caption{Parallel}
    \label{fig:dcn-parallel}
    \end{subfigure}
    \caption{Visualization of {DCN-V2}. $\otimes$ represents the cross operation in Eq. \eqref{eq:cross_layer}, \emph{i.e.}, $\vecx_{l+1} =\vecx_0 \odot (W_l \vecx_l +  \vecb_l) + \vecx_l$.}
    \label{fig:dcn-visualization}
\end{figure}

This section describes a novel model architecture --- {DCN-V2} --- to learn both explicit and implicit feature interactions. {DCN-V2} starts with an \emph{embedding layer}, followed by a \emph{cross network} containing multiple cross layers that models explicit feature interactions, and then combines with a \emph{deep network} that models implicit feature interactions. The improvements made in {DCN-V2} are {\bf critical for putting {DCN} into practice for highly-optimized production systems}. {DCN-V2} significantly improves the expressiveness of DCN \cite{wang2017deep} in modeling complex explicit cross terms in web-scale production data, while maintaining its elegant formula for easy deployment. The function class modeled by {DCN-V2} is a strict superset of that modeled by DCN. The overall model architecture is depicted in Fig. \ref{fig:dcn-visualization}, with two ways to combine the cross network with the deep network: (1) stacked and (2) parallel.  In addition, observing the low-rank nature of the cross layers, we propose to leverage a mixture of low-rank cross layers to achieve healthier trade-off between model performance and efficiency.

\subsection{Embedding Layer}

The embedding layer takes input as a combination of categorical (sparse) and dense features, and outputs $\vecx_0 \in \mathbb{R}^d$. For the $i$-th categorical feature, we project it from a high-dimensional sparse space to a lower-dimensional dense space via
	$\vecx_{\text{embed}, i} = W_{\text{embed},i} \vece_i$,
where $\vece_i \in \{0, 1\}^{v_i}$; $W \in \mathbb{R}^{e_i \times v_i}$ is a learned projection matrix; $\vecx_{\text{embed}, i} \in \mathbb{R}^{e_i}$ is the dense embedded vector; $v_i$ and $e_i$ represents vocab and embedding sizes respectively. For multivalent features, we use the mean of the embedded vectors as the final vector. 

The output is the concatenation of all the embedded vectors and the normalized dense features:
	$\vecx_0 = [\vecx_{\text{embed}, 1}; \ldots; \vecx_{\text{embed}, n}; x_{\text{dense}}]$.

Unlike many related works \cite{song2019autoint, lian2018xdeepfm, qu2016product, guo2017deepfm, naumov2019deep, he2017neural} which requires $e_i=e_j~\forall i, j$, our model accepts arbitrary embedding sizes. This is particularly important for industrial recommenders where the vocab size varies from $O(10)$ to $O(10^5)$. Moreover, our model isn't limited to the above described embedding method; any other embedding techniques such as hashing could be adopted. 

\subsection{Cross Network}
\begin{figure}[htbp]
  \centering
  \includegraphics[width=3in]{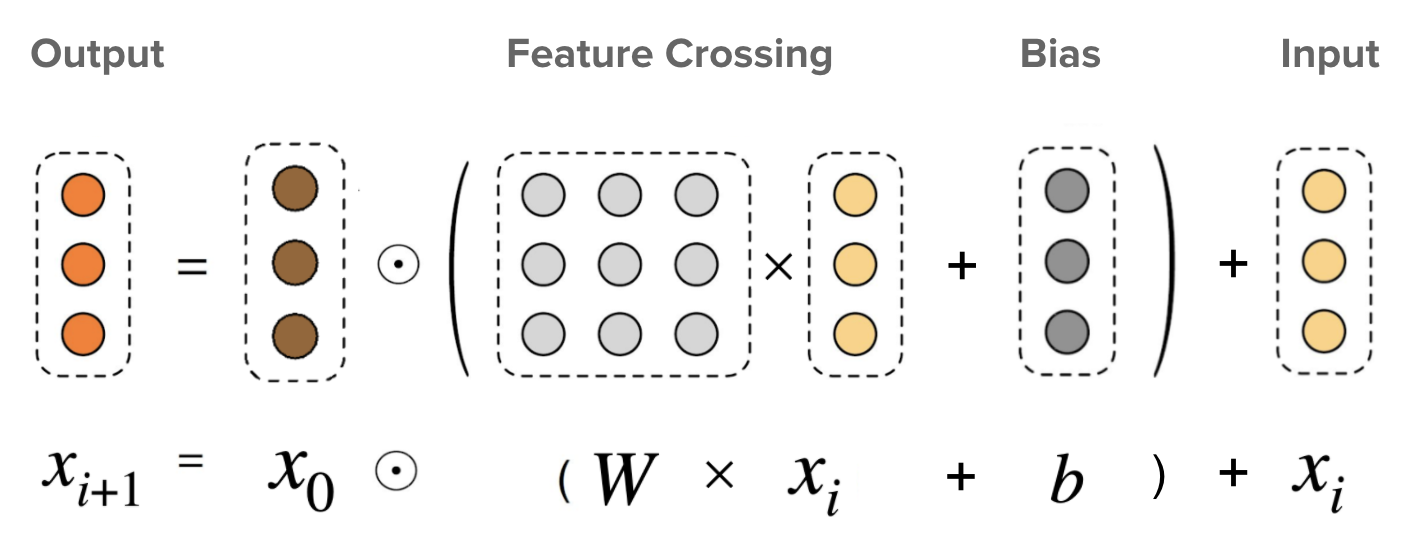}
\vspace{-2ex}
\caption{Visualization of a cross layer.}
  \label{fig:cross_sub_network}
\vspace{-1ex}
\end{figure}

The core of {DCN-V2} lies in the cross layers that create explicit feature crosses. Eq. \eqref{eq:cross_layer} shows the $(l+1)^\text{th}$ cross layer.
\begin{equation}
\label{eq:cross_layer}
	\vecx_{l+1} =\vecx_0 \odot (W_l \vecx_l +  \vecb_l) + \vecx_l
\end{equation}

where $\vecx_0 \in \mathbb{R}^{d}$ is the base layer that contains the original features of order 1, and is normally set as the embedding (input) layer. $\vecx_l,\vecx_{l+1} \in \mathbb{R}^{d}$, respectively, represents the input and output of the  $(l+1)$-th cross layer. $W_l \in \mathbb{R}^{d \times d}$ and $\vecb_l \in \mathbb{R}^{d}$ are the learned weight matrix and bias vector. \autoref{fig:cross_sub_network} shows how an individual cross layer functions.

For an $l$-layered cross network, the highest polynomial order is $l+1$ and the network contains all the feature crosses up to the highest order. Please see \autoref{sec:poly_analysis} for a detailed analysis, both from bitwise and feature-wise point of views. When $W = {\bf 1} \times \vecw^\top$, where ${\bf 1}$ represents a vector of ones, {DCN-V2} falls back to {DCN}.

The cross layers could only reproduce polynomial function classes of bounded degree; any other complex function space could only be approximated\footnote{Any function with certain smoothness assumptions can be well-approximated by polynomials. In fact, we've observed in our experiments that cross network alone was able to achieve similar performance as traditional deep networks.}. Hence, we introduce a deep network next to complement the modeling of the inherent distribution in the data.

\subsection{Deep Network}
The $l^\text{th}$ deep layer's formula is given by
	$\vech_{l+1} = f(W_l \vech_l +  \vecb_l)$,
where $\vech_l \in \mathbb{R}^{d_l}, \vech_{l+1} \in \mathbb{R}^{d_{l+1}}$, respectively, are the input and output of the $l$-th deep layer; $W_{l} \in \mathbb{R}^{d_{l} \times d_{l+1}}$ is the weight matrix and $\vecb_l \in \mathbb{R}^{d_{l+1}}$ is the bias vector; $f(\cdot)$ is an elementwise activation function and we set it to be ReLU; any other activation functions are also suitable.

\subsection{Deep and Cross Combination}
We seek structures to combine the cross network and deep network. Recent literature adopted two structures: stacked and parallel. In practice, we have found that which architecture works better is data dependent. Hence, we present both structures:

{\bf Stacked Structure (\autoref{fig:dcn-stack}):} The input $\vecx_0$ is fed to the cross network followed by the deep network, and the final layer is given by $\vecx_{\text{final}} = \vech_{L_d}, ~ \vech_{0} = \vecx_{L_c}$, which models the data as $f_{\text{deep}} \circ f_{\text{cross}}$.

{\bf Parallel Structure (\autoref{fig:dcn-parallel}):} The input $\vecx_0$ is fed in parallel to both the cross and deep networks; then, the outputs $\vecx_{L_c}$ and $\vech_{L_d}$ are concatenated to create the final output layer $\vecx_{\text{final}} = [\vecx_{L_c}; \vech_{L_d}]$. This structure models the data as $f_{\text{cross}} + f_{\text{deep}}$.
	

In the end, the prediction $\hat y_i$ is computed as:
	$\hat y_i = \sigma(\vecw_{\text{logit}}^\top\vecx_{\text{final} })$,
where $\vecw_{\text{logit}}$ is the weight vector for the logit, and $\sigma(x) = 1 / (1 + \exp(-x))$. For the final loss, we use the Log Loss that is commonly used for learning to rank systems especially with a binary label (e.g., click). Note that {DCN-V2} itself is both prediction-task and loss-function agnostic.
\begin{equation*}
	\begin{split}
	\text{loss} = &-\frac{1}{N} \sum_{i = 1}^N y_i \log(\hat y_i) + (1 - y_i) \log(1 - \hat y_i) + 
    \lambda \sum_l \|W_l\|^2_2,
    \end{split}
\end{equation*}
where $\hat y_i$'s are predictions; $y_i$'s are the true labels; $N$ is the total number of inputs; and $\lambda$ is the $L_2$ regularization parameter. 

\subsection{Cost-Effective Mixture of Low-Rank DCN}
In real production models, the model capacity is often constrained by limited serving resources and strict latency requirements. It is often the case that we have to seek methods to reduce cost while maintaining the accuracy. Low-rank techniques \cite{golub1996matrix} are widely used \cite{jaderberg2014speeding, yu2017compressing, chen2018adaptive, wang2019block, halko2011finding, drineas2005nystrom} to reduce the computational cost. It approximates a dense matrix $M \in \mathbb{R}^{d \times d}$ by two tall and skinny matrices $U, V \in \mathbb{R}^{d \times r}$. When $r \le d/2$, the cost will be reduced. 
However, they are most effective when the matrix shows a large gap in singular values or a fast spectrum decay. In many settings, we indeed observe that the learned matrix is numerically low-rank in practice.

\begin{figure}[htbp]
\centering
    \begin{subfigure}[b]{0.2\textwidth}  
    \includegraphics[width=\textwidth]{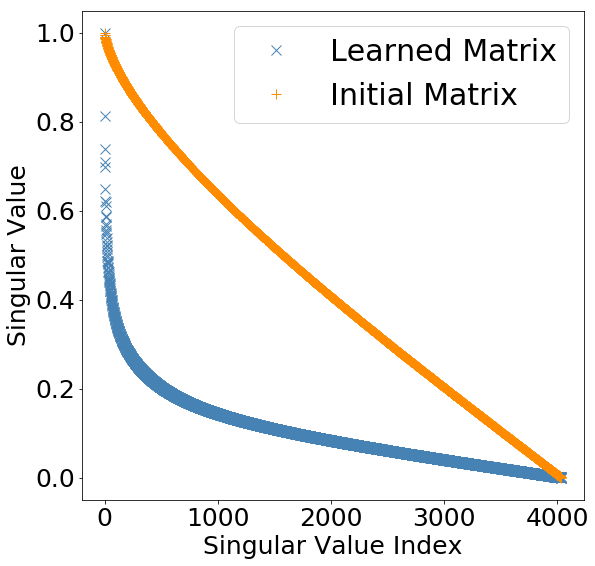}
    \caption{Singular Values}
    \label{fig:dcn-sval}
    \end{subfigure}
    \hfill
    \begin{subfigure}[b]{0.23\textwidth}  
    \includegraphics[width=\textwidth]{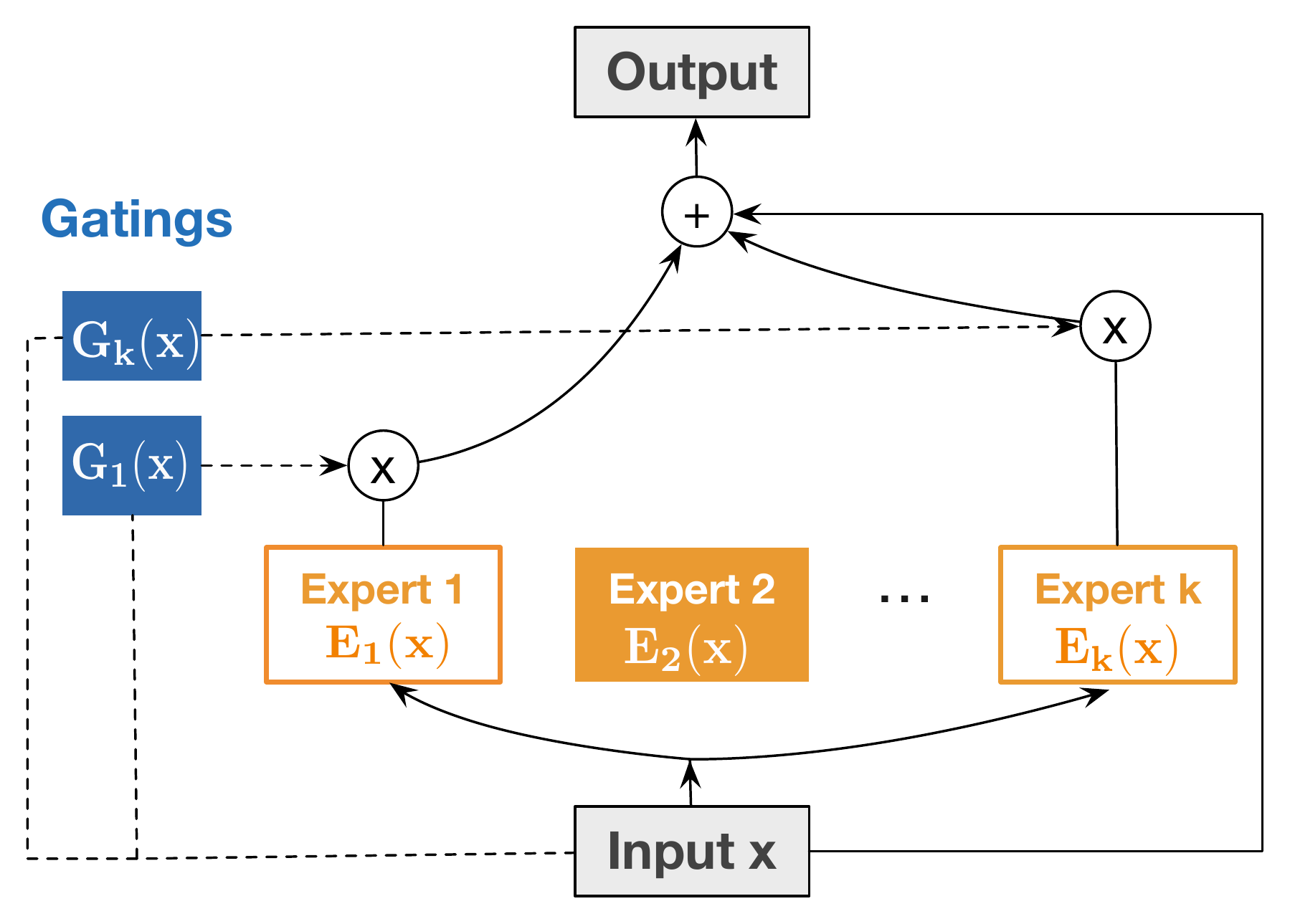}
    \caption{Mixture of Low-rank Experts}
    \label{fig:mixture-dcn}
    \end{subfigure}
    \caption{Left: Singular value decay of the learned {DCN-V2} weight matrix. The singular values are normalized and $1 = \sigma_1 \ge \sigma_2 \ge \ldots \ge \sigma_k$. {\color{orange}$+$} represents the randomly initialized truncated normal matrix; {\color{blue}$\times$} represents the final learned matrix. Right: Visualization of mixture of low-rank cross layer.}
\end{figure}

Fig. \ref{fig:dcn-sval} shows the singular decay pattern of the learned matrix $W$ in {DCN-V2} (see Eq. \eqref{eq:cross_layer}) from a production model. Compared to the initial matrix, the learned matrix shows a much faster spectrum decay pattern. Let's define the numerical rank $R_T$ with tolerance T to be $\text{argmin}_k (\sigma_k < T \cdot \sigma_1)$, where $\sigma_1 \ge \sigma_2 \ge, \ldots, \ge \sigma_n$ are the singular values. Then, $R_T$ means majority of the mass up to tolerance $T$, is preserved in the top $k$ singular values. In the field of machine learning and deep learning, a model could still work surprisingly well with a reasonably high tolerance $T$ \footnote{This is very different from the filed of scientific computing (\emph{e.g.}, solving linear systems), where the approximation accuracy need to be very high. For problems such as CTR prediction, some errors could introduce regularization effect to the model.}.

Hence, it is well-motivated to impose a low-rank structure on $W$. Eq \eqref{eq:low_rank_cross_layer} shows the resulting $(l+1)$-th low-rank cross layer
\begin{equation}
\label{eq:low_rank_cross_layer}
	\vecx_{l+1} = \vecx_0 \odot \Big(U_l \big(V_l^\top \vecx_i\big) + \vecb_l \Big) + \vecx_i
\end{equation}
where $U_l, V_l \in \mathbb{R}^{d \times r}$ and $r \ll d$. Eq \eqref{eq:low_rank_cross_layer} has two \emph{interpretations}: 1) we learn feature crosses in a subspace; 2) we project the input $\vecx$ to lower-dimensional $\mathbb{R}^r$, and then project it back to $\mathbb{R}^d$. The two interpretations have inspired the following two model improvements.

Interpretation 1 inspires us to adopt the idea from Mixture-of-Experts (MoE) \cite{shazeer2017outrageously, jacobs1991adaptive, eigen2013learning, ma2018modeling}. MoE-based models consist of two components: experts (typically a small network) and gating (a function of inputs). In our case, instead of relying on one single expert (Eq \eqref{eq:low_rank_cross_layer}) to learn feature crosses, we leverage multiple such experts, each learning feature interactions in a different subspaces, and adaptively combine the learned crosses using a gating mechanism that depends on input $\vecx$. The resulting mixture of low-rank cross layer formulation is shown in Eq. \eqref{eq:mixture_low_rank_cross_layer} and depicted in \autoref{fig:mixture-dcn}.
\begin{equation}
\label{eq:mixture_low_rank_cross_layer}
    \begin{split}
	\vecx_{l+1} &= \sum\nolimits_{i=1}^K G_i(\vecx_l) E_i(\vecx_l) + \vecx_l \\
	E_i(\vecx_l) &= \vecx_0 \odot \Big(U_l^i \big(V_l^{i\top}\vecx_l \big)  + \vecb_l \Big )
	\end{split}
\end{equation}
where $K$ is the number of experts; $G_i(\cdot): \mathbb{R}^d \mapsto \mathbb{R}$ is the gating function, common sigmoid or softmax; $E_i(\cdot): \mathbb{R}^d \mapsto \mathbb{R}^d$ is the $i^\text{th}$ expert in learning feature crosses. $G(\cdot)$ dynamically weights each expert for input $\vecx$, and when $G(\cdot) \equiv 1$, Eq \eqref{eq:mixture_low_rank_cross_layer} falls back to Eq \eqref{eq:low_rank_cross_layer}. 

Interpretation 2 inspires us to leverage the low-dimensional nature of the projected space. Instead of immediately projecting back from dimension $d'$ to $d$ ($d' \ll d$), we further apply nonlinear transformations in the projected space to refine the representation \cite{fan2019multiscale}.
\begin{equation}
\label{eq:mixture_low_rank_with_c_cross_layer}
	E_i(\vecx_l) = \vecx_0 \odot \Big(
	U_l^i \cdot g \big( C_l^i \cdot g\big(V_l^{i\top}\vecx_l \big) \big)  + \vecb_l 
	\Big )
\end{equation}
where $g(\cdot)$ represents any nonlinear activation function. 

{\bf Discussions.} This section aims to make effective use of the fixed memory/time budget to learn meaningful feature crosses. From Eqs \eqref{eq:cross_layer}--\eqref{eq:mixture_low_rank_with_c_cross_layer}, each formula represents a strictly larger function class assuming a fixed \#params. 

Different from many model compression techniques where the compression is conducted post-training, our model imposes the structure prior to training and jointly learn the associated parameters with the rest of the parameters. Due to that, the cross layer is an integral part of the nonlinear system $f(\vecx) = \big(f_k(W_k) \circ \cdots \circ f_1(W_1)\big) (\vecx)$, where $(f_{i+1} \circ f_{i}) (\cdot) \coloneqq f_{i+1}(f_i(\cdot))$. Hence, the training dynamics of the overall system might be affected, and it would be interesting to see how the global statistics, such as Jacobian and Hession matrices of $f(\vecx)$, are affected. We leave such investigations to future work.


\subsection{Complexity Analysis}
Let $d$ denote the embedding size, $L_c$ denote the number of cross layers, $K$ denote the number of low-rank DCN experts. Further, for simplicity, we assume each expert has the same smaller dimension $r$ (upper bound on the rank). 
The time and space complexity for the cross network is $O(d^2 L_c)$, and for mixture of low-rank DCN ({DCN-Mix}) it's efficient when $rK \ll d$ with $O(2drKL_c)$. 



\section{Model Analysis}
\label{sec:dcn_analysis}
This section analyzes {DCN-V2} from polynomial approximation point of view, and makes connections to related work. We adopt the notations from \cite{wang2017deep}.

{\bf Notations.} Let the embedding vector $\vecx = [\vecx_1; \vecx_2; \ldots; \vecx_k] = [x_1, x_2, \ldots, x_d] \in \mathbb{R}^d$ be a column vector, where $\vecx_i \in \mathbb{R}^{e_i}$ represents the $i$-th feature embedding, and $x_i$ represents the $i$-th element in $\vecx$. Let multi-index $\vecalpha = [\alpha_1, \cdots, \alpha_d] \in \mathbb{N}^d$ and $|\vecalpha| = \sum_{i=1}^d \alpha_i$. $C_a^b \coloneqq \bigl\{\vecy \in \{1,\cdots, a\}^b \mathrel{\big|} \forall i < j, y_i > y_j  \bigr\}$. Let $\mathbf{1}$ be a vector of all 1's, and $I$ be an identity matrix. We use capital letters for matrices, bold lower-case letters for vectors, and normal lower-case letters for scalars.

\subsection{Polynomial Approximation}
\label{sec:poly_analysis}
We analyze {DCN-V2} from two perspectives of polynomial approximation ---
1) Considering each element (bit) $x_i$ as a unit, and analyzes interactions among the elements (\autoref{thm:cross_x0_bitwise}); and 2) Considering each feature embedding $\vecx_i$ as a unit, and only analyzes the feature-wise interactions (\autoref{thm:cross_x0_featurewise} ) (proofs in Appendix). 

\begin{theorem}[Bitwise]
\label{thm:cross_x0_bitwise}
	Assume the input to an $l$-layer cross network be $\vecx \in \mathbb{R}^d$, the output be $f_l(\vecx)={\bf 1}^\top \vecx^l$, and the $i^\text{th}$ layer is defined as $\vecx^i =\vecx \odot W^{(i-1)}\vecx^{i-1} +\vecx^{i-1}$. Then, the multivariate polynomial $f_l(\vecx)$ reproduces polynomials in the following class:
    $$\biggl\{\sum_{\vecalpha} c_{\vecalpha}\left(W^{(1)}, \ldots, W^{(l)}\right) x_1^{\alpha_1}x_2^{\alpha_2}\ldots x_d^{\alpha_d} \mathrel{\bigg|}  0 \le |\vecalpha| \le l+1, \vecalpha \in \mathbb{N}^d \biggr\},$$
    where $c_\vecalpha = \sum_{\vecj \in C_l^{|\vecalpha|-1}}  \sum_{\veci \in P_\vecalpha} \prod_{k=1}^{|\vecalpha|-1} w_{i_k i_{k+1}}^{(j_k)}$, $w_{i j}^{(k)}$ is the $(i,j)^\text{th}$ element of matrix $W^{(k)}$, and $P_\vecalpha = \text{Permutations}~(\cup_i \{\underbrace{i, \ldots, i}_{\alpha_i \text{times}} \mathrel{|} \alpha_i \neq 0\})$.
\end{theorem}


\begin{theorem}[feature-wise]
\label{thm:cross_x0_featurewise}
  With the same setting as in \autoref{thm:cross_x0_bitwise}, we further assume input $\vecx = [\vecx_1; \ldots; \vecx_k]$ contains $k$ feature embeddings and consider each $\vecx_i$ as a unit. Then, the output $\vecx^l$ of an $l$-layer cross network creates all the feature interactions up to order $l+1$. Specifically, for features with their (repeated) indices in $I$, let $P_I = Permutations (I)$, then their order-$p$ interaction is given by:
\begin{equation*}
\begin{split}
\sum_{{\bf i} \in P_I} \sum_{\vecj \in C_p^{p-1}} \vecx_{i_1} \odot \left(W_{i_1, i_2}^{(j_1)} \vecx_{i_2} \odot \ldots \odot \left(W_{i_k, i_{k+1}}^{(j_{k})} \vecx_{i_{l+1}}\right) \right)
\end{split}
\end{equation*}

\end{theorem}


From both bitwise and feature-wise perspectives, the cross network is able to create all the feature interactions up to order $l+1$ for an $l$-layered cross network. Compared to DCN-V, {DCN-V2} characterizes the same polynomial class with more parameters and is more expressive. Moreover, the feature interactions in {DCN-V2} is more expressive and can be viewed both bitwise and feature-wise, whereas in DCN it is only bitwise \cite{wang2017deep, lian2018xdeepfm, song2019autoint}.

\subsection{Connections to Related Work}
We study the connections between {DCN-V2} and other SOTA feature interaction learning methods; we only focus on the feature interaction component of each model and ignore the DNN component. For comparison purposes, we assume the feature embeddings are of equal size $e$.

{\bf DCN.} Our proposed model was largely inspired from DCN \cite{wang2017deep}. Let's take the efficient projection view of DCN \cite{wang2017deep}, \emph{i.e.}, it implicitly generates all the pairwise crosses and then projects it to a lower-dimensional space; {DCN-V2} is similar with a different projection structure.
    \begin{equation*}
    	\begin{split}
    	\vecx_{\text{DCN}}^\top = 
        \vecx_{\text{pairs}}
        \left[\begin{smallmatrix}
        	\vecw & \veczero & \ldots & \veczero \\
            \veczero & \vecw & \ldots & \veczero \vspace{-1ex}\\
            \vdots   & \vdots& \ddots & \vdots \\
            \veczero & \veczero & \ldots & \vecw
        \end{smallmatrix}\right], 
        \vecx_{\text{{DCN-V2}}}^\top = 
        \vecx_{\text{pairs}}
         \left[\begin{smallmatrix}
        	\vecw_1 & \veczero & \ldots & \veczero \\
            \veczero & \vecw_2 & \ldots & \veczero \vspace{-1ex}\\
            \vdots   & \vdots& \ddots & \vdots \\
            \veczero & \veczero & \ldots & \vecw_d
        \end{smallmatrix}\right]
        \end{split}
    \end{equation*}
where
$\vecx_{\text{pairs}} = [x_i \tilde x_j]_{\forall i, j}$
contains all the $d^2$ pairwise interactions between $\vecx_0$ and $\tilde \vecx$; $\vecw \in \mathbb{R}^d$ is the weight vector in DCN-V; $\vecw_i \in \mathbb{R}^d$ is the $i^\text{th}$ column of the weight matrix in {DCN-V2} (Eq.\eqref{eq:cross_layer}).

{\bf DLRM and DeepFM.} Both are essentially 2nd-order FM without the DNN component (ignoring small differences). Hence, we simplify our analysis and compare with FM which has formula
	$\vecx^\top \vecbeta + \sum_{i<j} w_{ij} \langle \vecx_i, \vecx_j \rangle$.
This is equivalent to 1-layer {DCN-V2} (Eq. \eqref{eq:cross_layer} without residual term) with a structured weight matrix.
\begin{equation*}
\label{eq:dcn-fm}
\begin{split}
    \mathbf{1}^\top \left(
    \left[\begin{smallmatrix}
        \vecx_1 \\
        \vecx_2 \vspace{-1ex}\\
        \vdots  \\
        \vecx_k
    \end{smallmatrix}\right]
    \odot
    \left(
    \left[\begin{smallmatrix}
        \veczero & w_{12} I & \cdots & w_{1k}I \\
        \veczero & \veczero & \cdots & w_{2k} I \vspace{-1ex}\\
        \vdots  & \vdots  & \ddots & \vdots  \\
        \veczero & \veczero & \cdots & \veczero
    \end{smallmatrix}\right]
    \left[\begin{smallmatrix}
        \vecx_1 \\
        \vecx_2 \vspace{-1ex}\\
        \vdots  \\
        \vecx_k
    \end{smallmatrix}\right]
    +
    \left[\begin{smallmatrix}
        \vecbeta_1 \\
        \vecbeta_2 \vspace{-1ex}\\
        \vdots  \\
        \vecbeta_k
    \end{smallmatrix}\right]
    \right)
    \right)
\end{split}
\end{equation*}

{\bf xDeepFM.} The $h$-th feature map at the $k$-th layer is given by:
\begin{equation*}
\label{eq:xdeepfm}
    \vecx_{h, *}^k = \sum\nolimits_{i=1}^{k-1} \sum\nolimits_{j=1}^m w_{ij}^{k, h} (\vecx_{i, *}^{k-1} \odot \vecx_j)
\end{equation*}
The $h$-th feature map at the 1st layer is equivalent to 1-layer {DCN-V2} (Eq. \eqref{eq:cross_layer} without residual term).
\begin{equation*}
\label{eq:dcn-fm2}
  \vecx_{h, *}^1 = [I, I, \cdots, I]\left(\vecx \odot (W\vecx)\right) = \sum\nolimits_{i=1}^k \vecx_i \odot (W_{i,:} \vecx)
\end{equation*}
where the $(i,j)$-th block $W_{i,j} = w_{ij} \cdot I$, and $W_{i,:} \coloneqq [W_{i,1}, \ldots, W_{i,k}]$. 

{\bf AutoInt.} The interaction layer of AutoInt adopted the multi-head self-attention mechanism. For simplicity, we assume a single head is used in AutoInt; multi-head case could be compared summarily using concatenated cross layers.  

From a high-level view, the 1st layer of AutoInt outputs $\widetilde \vecx = [\widetilde \vecx_1; \widetilde \vecx_2; \ldots; \widetilde \vecx_k]$, where $\widetilde \vecx_i$ encodes all the 2nd-order feature interactions with the i-th feature. Then, $\widetilde \vecx$ is fed to the 2nd layer to learn higher-order interactions. This is the same as {DCN-V2}. 

From a low-level view (ignoring the residual terms),
\begin{equation*}
\small
    \label{eq:autoint}
    \begin{split}
        \widetilde \vecx_i
        &= ReLU\left(\sum\nolimits_{j=1}^k \frac{\exp \left(\langle W_{\text{q}} \vecx_i, W_{\text{k}} \vecx_j \rangle \right)}{\sum\nolimits_j \exp \left(\langle W_{\text{q}} \vecx_i, W_{\text{k}} \vecx_j \rangle \right)} (W_{\text{v}} \vecx_j)\right) \\
        &= ReLU\big(\sum\nolimits_{j=1}^k \text{softmax}( \vecx_i^\top \widetilde W \vecx_j)~ W_{\text{v}} \vecx_j \big)
\end{split}
\end{equation*}
where $\langle \cdot, \cdot \rangle$ represents inner (dot) product, and $\widetilde W = W_{\text{q}} W_{\text{k}}$.
While in {DCN-V2}, 
\begin{equation}
    \label{eq:autoint_dcn}
    \begin{split}
        \widetilde \vecx_i = \sum\nolimits_{j=1}^k \vecx_i \odot (W_{i,j} \vecx_j) = \vecx_i \odot (W_{i,:} \vecx)
    \end{split}
\end{equation}
where $W_{i,j}$ represents the $(i,j)$-th block of $W$. It is clear that the difference lies in how we model the feature interactions. AutoInt claims the non-linearity was from ReLU($\cdot$); we consider each summation term to also contribute. Differently, {DCN-V2} used $\vecx_i \odot W_{i,j} \vecx_j$. 

{\bf PNN.} The inner-product version (IPNN) is similar to FM. For the outer-product version (OPNN), it first explicitly creates all the $d^2$ pairwise interactions,
and then projects them to a lower dimensional space $d'$ using a $d'$ by $d^2$ dense matrix. Differently, {DCN-V2} implicitly creates the interactions using a structured matrix.

\section{Research Questions}
\label{sec:research_qs}
We are interested to seek answers for these following research questions:

\begin{itemize}[leftmargin=2.5em]
    \item[\bf RQ1]  When would feature interaction learning methods become more efficient than ReLU-based DNNs? 
    \item[\bf RQ2]  How does the feature-interaction component of each baseline perform without integrating with DNN? 
    \item[\bf RQ3]  How does the proposed mDCN approaches compare to the baselines? Could we achieve healthier trade-off between model accuracy and cost through mDCN and the mixture of low-rank DCN?
    \item[\bf RQ4] How does the settings in mDCN affect model quality?
    \item[\bf RQ5] Is mDCN capturing important feature crosses? Does the model provide good understandability?
\end{itemize}

Throughout the paper, ``CrossNet" or ``CN" represents the cross network; suffix ``Mix" denotes the mixture of low-rank version. 

\section{Empirical understanding of feature crossing techniques (RQ1)}
\label{sec:exp_synthetic}
Many recent works \cite{wang2017deep, cheng2016wide, guo2017deepfm, beutel2018latent, qu2016product, lian2018xdeepfm, naumov2019deep} proposed to model explicit feature crosses that couldn't be learned efficiently from traditional neural networks. However, most works only studied public datasets with unknown cross patterns and noisy data; few work has studied in a clean setting with known ground-truth models. 
Hence, it's important to understand : 1) in which cases would traditional neural nets become inefficient; 2) the role of each component in the cross network of {DCN-V2}.

We use the cross network in DCN models to represent those feature cross methods and compare with ReLUs, which are commonly used in industrial recommender systems.
To simplify experiments and ease understanding, we assume each feature $x_i$ is of dimension one, and monomial $x_1^{\alpha_1}x_2^{\alpha_2}\cdots x_d^{\alpha_d}$ represents a $|\vecalpha|$-order interaction between features. 

{\bf Performance with increasing difficulty.}
Consider only 2nd-order feature crosses and let the ground-truth model be $f(\vecx) = \sum_{|\vecalpha|=2} w_{\vecalpha} x_1^{\alpha_1}x_2^{\alpha_2}\ldots x_d^{\alpha_d}$.
Then, the difficulty of learning $f(\vecx)$ depends on: 1) sparsity ($w_{\vecalpha}=0$), the number of crosses, and 2) similarity of the cross patterns (characterized by $\Var(w_{\vecalpha})$), meaning a change in one feature would simultaneously affect most feature crosses by similar amount. We create synthetic datasets with increasing difficulty in Eq. \eqref{eq:2nd-order}.
\begin{equation}
    \label{eq:2nd-order}
    \begin{split}
    f_1(\vecx) &= x_1^2+x_1x_2+x_3x_1+x_4x_1 \\
    f_2(\vecx) &= x_1^2+0.1x_1x_2+x_2x_3+0.1x_3^2 \\
    f_3(\vecx) &= \sum\nolimits_{(i,j) \in S} w_{ij} x_i x_j, ~~\vecx \in \mathbb{R}^{100}, |S|=100
    \end{split}
\end{equation}
where set $S$ and weights $w_{ij}$ are randomly assigned, and $x_i$'s are uniformly sampled from interval [-1, 1].

\autoref{tab:2nd-order} reports mean RMSE out of 5 runs and the model size. 
When the cross patterns are simple ($f_1$), both {DCN-V2} and DCN are efficient. When the patterns become more complicated ($f_3$), {DCN-V2} remains accurate while DCN degrades. DNN's performance remains poor even with a wider and deeper structure (layer sizes [200, 200] for $f_1$ and $f_2$, [1024, 512, 256] for $f_3$). This suggests the inefficiency of DNN in modeling monomial patterns. 

\begin{table}[htpb]
\footnotesize
\vspace{-5pt}
\caption{RMSE and Model Size (\# Parameters) for Polynomial Fitting of Increasing Difficulty.}
\vspace{-3.5ex}
\label{tab:2nd-order}
\begin{center}
\begin{tabular}{c|ll|ll|ll|ll}
\toprule
& \multicolumn{2}{c|}{DCN (1Layer)}  & \multicolumn{2}{c|}{DCN-V2 (1Layer)}  & \multicolumn{2}{c|}{DNN (1Layer)} & \multicolumn{2}{c}{DNN (large)}\\
& RMSE & Size & RMSE & Size & RMSE & Size & RMSE & Size \\
\midrule
$f_1$& 8.9E-13 & 12 & {\bf 5.1E-13} & 24 & 2.7E-2 & 24  & 4.7E-3 & 41K 
\\
$f_2$ & 1.0E-01 & 9  & {\bf 4.5E-15} & 15 & 3.0E-2 & 15 & 1.4E-3 & 41K\\
$f_3$ & 2.6E+00 & 300 & {\bf 6.7E-07} & 10K & 2.7E-1 & 10K & 7.8E-2 & 758K\\
\bottomrule
\end{tabular}
\end{center}
\vspace{-5pt}
\end{table}

{\bf Role of each component.}
We also conducted ablation studies on homogeneous polynomials of order 3 and 4, respectively. For each order, we randomly selected 20 cross terms from $\vecx \in \mathbb{R}^{50}$.

\begin{figure}[htbp]
\centering
    \begin{subfigure}[b]{0.22\textwidth}  
    \includegraphics[width=\textwidth]{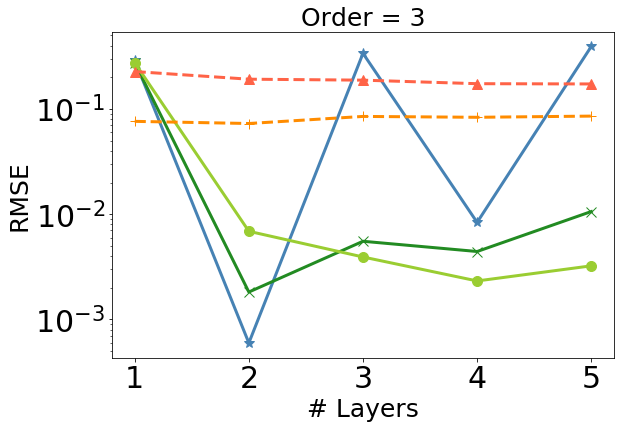}
    \end{subfigure}
    \hfill
    \begin{subfigure}[b]{0.208\textwidth}  
    \includegraphics[width=\textwidth]{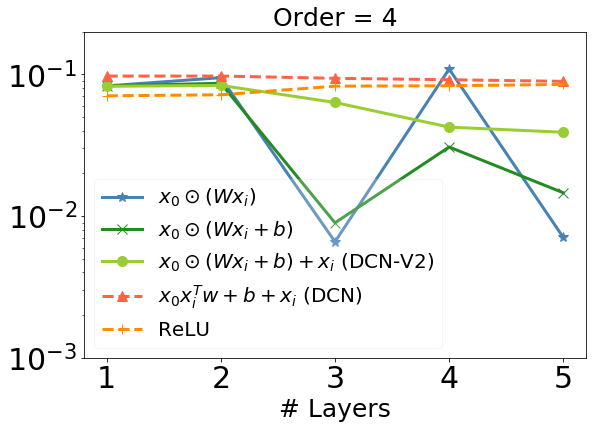}
    \end{subfigure}
    \vspace{-1ex}
    \caption{Homogeneous polynomial fitting of order 3 and 4. $x$-axis represents the number of layers used; $y$-axis represents RMSE (the lower the better). In the legend, the top 3 models are {DCN-V2} with different component(s) included.}
\label{fig:dcn-poly-order}
\end{figure}

\autoref{fig:dcn-poly-order} shows the change in mean RMSE with layer depth. Clearly, $\vecx_0 \odot (W\vecx_i)$ models order-$d$ crosses at layer $d$-1, which is verified by that the best performance for order-3 polynomial is achieved at layer 2 (similar for order-4). At other layers, however, the performance significantly degrades. This is where the bias and residual terms are helpful --- they create and maintain all the crosses up to the highest order. This reduces the performance gap between layers, and stabilizes the model when redundant crosses are introduced. This is particularly important for real-world applications with unknown cross patterns.

Fig. \ref{fig:dcn-poly-order} also reveals the limited expressiveness of {DCN} in modeling complicated cross patterns.

{\bf Performance with increasing layer depth.} We now study scenarios closer to real-world settings, where the cross terms are of a combined order.  
\begin{equation*}
\small
    \label{eq:combined-order}
    \begin{split}
    f(\vecx) = & \vecx^\top \vecw + 
\sum_{\vecalpha \in S} w_{\vecalpha} x_1^{\alpha_1}x_2^{\alpha_2}\cdots x_d^{\alpha_d}  +
0.1 \sin(2\vecx^\top \vecw_{s} + 0.1) +
0.01 \epsilon
    \end{split}
\end{equation*}
where the randomly chosen set $S = S_2 \cup S_3 \cup S_4$, $|S_2|=20, |S_3|=10, |S_4|=5$, and $\forall \vecalpha \in S_i, |\vecalpha|=i$; sine introduces perturbations and $\epsilon$ represents Gaussian noises.

\autoref{tab:combined-order} reports the mean RMSE out of 5 runs. With the increase of layer depth, CN-M was able to capture higher-order feature crosses in the data, resulting in improved performance. Thanks to the bias and residual terms, the performance didn't degrade beyond layer 3, where redundant feature interactions were introduced. 

\begin{table}[htbp]
\small
\caption{Combined-order (1 - 4) Polynomial Fitting.}
\vspace{-3.ex}
\label{tab:combined-order}
    \begin{subtable}[h]{0.45\textwidth}
        \centering
\begin{tabular}{c|ccccc}
\toprule
{\bf \#Layers} & 1 & 2 & 3 & 4 & 5\\
\midrule
{DCN-V2} & 1.43E-01 & 2.89E-02 & \bf 9.82E-03 & 9.87E-03 & 9.92E-03 \\
DNN& 1.32E-01 & 1.03E-01 & 1.03E-01 &  1.09E-01 & 1.05E-01\\
\bottomrule
\end{tabular}
    \end{subtable}
\end{table}

To summarize, ReLUs are inefficient in capturing explicit feature crosses (multiplicative relations) even with a deeper and larger network. This is well aligned with previous studies \cite{beutel2018latent}. 
The accuracy considerably degrades when the cross patterns become more complicated. DCN accurately captures simple cross patterns but fails at more complicated ones. {DCN-V2}, on the other hand, remains accurate and efficient for complicated cross patterns.

\section{Experimental Results (RQ2 - RQ5)}
\label{sec:exp_public}
This section empirically verifies the effectiveness of {DCN-V2} in feature interaction learning across 3 datasets and 2 platforms, compared with SOTA. In light of recent concerns about poor reproducibility of published results \cite{dacrema2019we, musgrave2020metric, rendle2020neural}, we conducted a fair and comprehensive experimental study with extensive hyper-parameter search to properly tune all the baselines and proposed approaches. In addition, for each optimal setup, we train 5 models with different random initialization, and report the mean and standard deviation. 

\autoref{sec:performance_feature_interaction} studies the performance of the feature-cross learning components (\textbf{RQ2}) between baselines \emph{without} integrating with DNN ReLU layers (similar to \cite{lian2018xdeepfm, song2019autoint}); only sparse features are considered for a clean comparison. \autoref{sec:performance_baselines} compares {DCN-V2} with all the baselines comprehensively (\textbf{RQ3}). \autoref{sec:hyper-parameters} evaluates the influence of hyper-parameters on the performance of {DCN-V2} (\textbf{RQ4}). \autoref{sec:model_understanding} focuses on model understanding (\textbf{RQ5}) of whether we are indeed discovering meaningful feature crosses with {DCN-V2}.

\subsection{Experiment Setup}
\label{sec:experiment_setup}
This section describes the experiment setup, including training datasets, baseline approaches, and details of the hyper-parameter search and training process.
\subsubsection{Datasets}
\autoref{tab:datasets} lists the statistics of each dataset:

\begin{table}[htpb]
\small
\caption{Datasets.}
\vspace{-3.5ex}
\label{tab:datasets}
\begin{center}
\begin{tabular}{c|ccc}
\toprule
{\bf Data} & \# Examples & \# Features & Vocab Size \\
\midrule
Criteo & 45M & 39 & 2.3M\\
MovieLen-1M &  740k & 7 & 3.5k\\
Production & $>$ 100B & NA & NA\\
\bottomrule
\end{tabular}
\end{center}
\end{table}

{\bf Criteo\footnote{http://labs.criteo.com/2014/02/kaggle-display-advertising-challenge-dataset}.} The most popular click-through-rate (CTR) prediction benchmark dataset contains user logs over a period of 7 days. We follow \cite{wang2017deep, song2019autoint} and use first 6 days for training, and randomly split the last day's data into validation and test set equally. We log-normalize ($\log(x+4)$ for feature-2 and $\log(x+1)$ for others) the 13 continuous features and embed the remaining 26 categorical features.

{\bf MovieLen-1M\footnote{https://grouplens.org/datasets/movielens}.} The most popular dataset for recommendation systems research. Each training example includes a {\ttfamily{$\langle$\small user-features, \small movie-features, \small rating$\rangle$}} triplet. Similar to AutoInt~\cite{song2019autoint}, we formalize the task as a regression problem. All the ratings for 1s and 2s are normalized to be 0s; 4s and 5s to be 1s; and rating 3s are removed. 6 non-multivalent categorical features are used and embedded. The data is randomly split into 80\% for training, 10\% for validation and 10\% for testing.

\subsubsection{Baselines.}
\label{sec:baselines}
We compare our proposed approaches with 6 SOTA feature interaction learning algorithms. A brief comparison between the approaches is highlighted in \autoref{tab:model_comparisons}.

\begin{table}[htpb]
\small
\caption{High-level comparison between models. Assuming the input $\vecx_0 = [\vecv_1; \ldots; \vecv_k]$ contains $k$ feature embeddings that each represented as $\vecv_i$. $\oplus$ denotes concatenation; $\otimes$ denotes outer-product; $\odot$ denotes Hadamard-product. $f_i(\cdot)$ represents implicit feature interactions, \emph{i.e.,} ReLU layers. In the last column, the `+' sign is on the logit level.}
\vspace{-2ex}
\label{tab:model_comparisons}
\begin{center}
\begin{tabular}{l|p{5mm}l|l}
\toprule
\multirow{ 2}{*}{{\bf Model}}  & \multicolumn{2}{c|}{ Explicit Interactions ($f_e$)} &  \multirow{ 2}{*}{\begin{tabular}{@{}l@{}}Final \\ Objective  \end{tabular}}  \\
\cmidrule(){2-3}
& Order & \multicolumn{1}{c|}{(Simplified) Key Formula}  & \\
\midrule
\multirow{ 2}{*}{{PNN \cite{qu2016product}}}  &\multirow{ 2}{*}{2}& $\vecx_o = [\vecv_i^\top \vecv_j \mathrel{|} \forall i, j]$ (IPNN) & \multirow{ 2}{*}{$f_i \circ f_e$}   \\
&&$\vecx_o = [\text{vec}(\vecv_i \otimes \vecv_j) \mathrel{|} \forall i, j]$ (OPNN) & \\
DeepFM \cite{guo2017deepfm} & $2$ & $\vecx_o = [\vecv_i^\top \vecv_j \mathrel{|} \forall i, j]$ &  $f_i + f_e$ \\
DLRM \cite{naumov2019deep} & 2 & $\vecx_o = [\vecv_i^\top \vecv_j \mathrel{|} \forall i, j]$& $f_i \circ f_e$\\
DCN \cite{wang2017deep} & $\ge 2$ & $\vecx_{i+1} = \vecx_0 \otimes \vecx_i \vecw_i$ &$f_i + f_e$ \\
xDeepFM \cite{lian2018xdeepfm} & $\ge 2$ & $\vecv_{h}^k = \sum_{i, j} w_{ij}^{kh} (\vecv_{i}^{k-1} \odot \vecv_j)$  & $f_i + f_e$ \\
AutoInt \cite{song2019autoint} & NA & $ \widetilde \vecv_i = g \left(\frac{\sum_{j}\exp (\langle W_q \vecv_i, W_k \vecv_j \rangle )W_v \vecv_j}{\sum_j \exp \left(\langle W_q \vecv_i, W_k \vecv_j \rangle \right)})\right)$ & $f_i + f_e$\\
\midrule
\multirow{ 2}{*}{{DCN-V2} (ours)} & \multirow{ 2}{*}{ $\ge 2$} & \multirow{ 2}{*}{$\vecx_i = \vecx_0 \odot (W_i \vecx_i)$} & $f_i \circ f_e$\\
&&&$f_i + f_e$ \\
\bottomrule
\end{tabular}
\end{center}
\end{table}

\subsubsection{Implementation Details.}
\label{sec:implementation_details}
All the baselines and our approaches are implemented in TensorFlow v1. For a fair comparison, all the implementations were identical across all the models except for the feature interaction component \footnote{We adopted implementation from \url{https://github.com/Leavingseason/xDeepFM}, \url{https://github.com/facebookresearch/dlrm} and \url{https://github.com/shenweichen/DeepCTR}}.

{\bf Embeddings.} All the baselines require each feature's embedding size to be the same except for DNN and DCN. Hence, we fixed it to be $ \text{Avg}\big(\sum_{\text{vocab}} 6\cdot(\text{vocab cardinality})^{\frac{1}{4}}\big)$ (39 for Criteo and 30 for Movielen-1M) for all the models\footnote{This formula is a rule-of-thumb number that is widely used \cite{wang2017deep}, also see \url{https://developers.googleblog.com/2017/11/introducing-tensorflow-feature-columns.html}}. 

{\bf Optimization.} We used Adam \cite{kingma2014adam} with a batch size of $512$ (128 for MovieLen). The kernels were initialized with He Normal \cite{he2015delving}, 
and biases to $\veczero$; the gradient clipping norm was 10; an exponential moving average with decay 0.9999 to trained parameters was applied.

{\bf Reproducibility and fair comparisons: hyper-parameters tuning and results reporting.} For all the baselines, we conducted a coarse-level (larger-range) grid search over the hyper-parameters, followed by a finer-level (smaller-range) search. To ensure reproducibility and mitigate model variance, for each approach and dataset, we report the mean and stddev out of 5 independent runs for the best configuration. We describe detailed settings below for Criteo; and follow a similar process for MovieLens with different ranges.

We first describe the hyper-parameters shared across the baselines. The learning rate was tuned from $10^{-4}$ to $10^{-1}$ on a log scale and then narrowed down to $10^{-4}$ to $5 \times 10^{-4}$ on a linear scale. The training steps were searched over \{150k, 160k, 200k, 250k, 300k\}. The number of hidden layers ranged in \{1, 2, 3, 4\} with their layer sizes in \{562, 768, 1024\}. And the regularization parameter $\lambda$ was in \{0, $3 \times 10^{-5}$, $10^{-4}$\}.

We then describe each model's own hyper-parameters, where the search space is designed based on reported setting. For DCN, the number of cross layers ranged from 1 to 4. For AutoInt, the number of attention layers was from 2 to 4; the attention embedding size was in \{20, 32, 40\}; the number of attention head was from 2 to 3; and the residual connection was either on or off. For xDeepFM, the CIN layer size was in \{100, 200\}, depth in \{2, 3, 4\}, activation was identity, computation was either direct or indirect. For DLRM, the bottom MLP layer sizes and numbers was in \{(512,256,64), (256,64)\}. For PNN, we ran for IPNN, OPNN and PNN*, and for the latter two, the kernel type ranged in \{full matrix, vector, number\}.
For all the models, the total number of parameters was capped at $1024^2 \times 5$ to limit the search space and avoid overly expensive computations.

\subsection{Performance of Feature Interaction Component Alone (RQ2)}
\label{sec:performance_feature_interaction}
We consider the feature interaction component alone of each model {\bf without their DNN component}. Moreover, we only consider the categorical features, as the dense features were processed differently among baselines. \autoref{tab:cross-only} shows the results on Criteo dataset. Each baseline was tuned similarly as in \autoref{sec:implementation_details}. There are two major observations. 1). Higher-order methods demonstrate a superior performance over 2nd-order methods. This suggests high-order crosses are meaningful in this dataset. 2). Among the high-order methods, cross network achieved the best performance and was on-par or slightly better compared to DNN.

\begin{table}[htpb]
\small
\caption{LogLoss (test) of feature interaction component of each model (no DNN). Only categorical features were used. In the `Setting' column, $l$ stands for number of layers.}
\vspace{-2ex}
\label{tab:cross-only}
\begin{center}
\begin{tabular}{c|l|cl}
\toprule
& Model & LogLoss & \multicolumn{1}{c}{Best Setting}\\
\midrule
\multirow{ 3}{*}{2nd} 
&PNN \cite{qu2016product} &0.4715 $\pm$ 4.430e-04 & OPNN, kernel=matrix\\
&FM      & 0.4736 $\pm$ 3.04E-04& \multicolumn{1}{c}{--} \\
\midrule
\multirow{ 5}{*}{$>$2} 
&CIN \cite{lian2018xdeepfm}  & 0.4719 $\pm$ 9.41E-04& l=3, cinLayerSize=100\\
&AutoInt \cite{song2019autoint}   & 0.4711 $\pm$ 1.62E-04& l=2, head=3, attEmbed=40\\
&DNN        & 0.4704 $\pm$ 1.57E-04& l=2, size=1024\\
\cline{2-4}
&CrossNet     & 0.4702 $\pm$ 3.80E-04& l=2\\
&CrossNet-Mix & \bf 0.4694 $\pm$ 4.35E-04 & l=5, expert=4, gate=$\frac{1}{1+e^{-x}}$\\

\bottomrule
\end{tabular}
\end{center}
\end{table}

\subsection{Performance of Baselines (RQ3)}
\label{sec:performance_baselines}
This section compares the performance between {DCN-V2} approaches and the baselines in an end-to-end fashion. Note that the best setting reported for each model was searched over a {\bf wide-ranged model capacity and hyper-parameter space} including the baselines. And if two settings performed on-par, we report the {\bf lower-cost} one. \autoref{tab:baseline_comparison} shows the best LogLoss and AUC (Area Under the ROC Curve) on testset for Criteo and MovieLen. For Criteo, a {\bf 0.001-level improvement} is considered significant (see \cite{song2019autoint, wang2017deep, guo2017deepfm}).  
We see that {DCN-V2} consistently outperformed the baselines (including DNN) and achieved a healthy quality/cost trade-off. It's also worth mentioning that the baselines' performances reported in \autoref{tab:baseline_comparison} were improved over the numbers reported by previous papers (see \autoref{tab:metrics_in_papers} in Appendix); however, when integrated with DNN, their performance gaps are closing up (compared to \autoref{tab:cross-only}) with their performances on-par and sometimes worse than the ReLU-based DNN with fine-granular model tuning.

{\bf Best Settings.} The optimal hyper-parameters are in \autoref{tab:baseline_comparison}. For {DCN-V2} models, both the `stacked' and `parallel' structures outperformed all the baselines, while `stacked' worked better on Criteo and `parallel' worked better on Movielen-1M. On Criteo, the setting was gate as constant, hard\_tanh activation for {DCN-Mix}; gate as softmax and identity activation for CrossNet. The best training steps was 150k for all the baselines; learning rate varies for all the models.


{\bf Model Quality --- Comparisons among baselines.} When integrating the feature cross learning component with a DNN, the advantage of higher-order methods is less pronounced, and the performance gap among all the models are closing up on Criteo (compared to \autoref{tab:cross-only}). \textbf{This suggests the importance of implicit feature interactions and the power of a well-tuned DNN model.}

For 2nd-order methods, DLRM performed inferiorly to DeepFM although they are both derived from FM. This might be due to DLRM's omission of the 1st-order sparse features after the dot-product layer. PNN models 2nd-order crosses more expressively and delivered better performance on MovieLen-1M; however on Criteo, its mean LogLoss was driven up by its high standard deviation. For higher-order methods, xDeepFM, AutoInt and DCN behaved similarly on Criteo, while on MovieLens xDeepFm showed a high variance.

{DCN-V2} achieved the best performance (0.001 considered to be significant on Criteo \cite{wang2017deep, lian2018xdeepfm, song2019autoint}) by explicitly modeling up to 3rd-order crosses beyond those implicit ones from DNN. {DCN-Mix}, the mixture of low-rank DCN, efficiently utilized the memory and reduced the cost by 30\% while maintaining the accuracy. Interestingly, CrossNet alone outperformed DNN on both datasets; we defer more discussions to \autoref{sec:crossnet_performance}.

{\bf Model Quality --- Comparisons with DNN.} DNNs are universal approximators and are tough-to-beat baselines when highly-optimized. Hence, we finely tuned DNN along with all the baselines, and used a larger layer size than those used in literature (\emph{e.g.}, 200 - 400 in \cite{lian2018xdeepfm, song2019autoint}). \textbf{To our surprise, DNN performed neck to neck with most baselines and even outperformed certain models.} 

Our hypothesis is that those explicit feature crosses from baselines were not modeled in an {\bf expressive} and {\bf easy-to-optimize} manner. The former makes its performanc easy to be matched by a DNN with large capacity. The latter would easily lead to trainability issues, making the model unstable, hard to identify a good local optima or to generalize. Hence, when integrated with DNN, the overall performance is dominated by the DNN component. This becomes especially true with a large-capacity DNN, which could already approximate some simple cross patterns.

In terms of expressiveness, consider the 2nd-order methods. PNN models crosses more expressively than DeepFM and DLRM, which resulted in its superior performance on MovieLen-1M. This also explains the inferior performance of DCN compared to {DCN-V2}.

In terms of trainability, certain models might be inherently more difficult to train and resulted in unsatisfying performance. Consider PNN. On MoiveLen-1M, it outperformed DNN, suggesting the effectiveness of those 2nd-order crosses. On Criteo, however, PNN's advantage has diminished and the averaged performance was on-par with DNN. This was caused by the instability of PNN. Although its best run was better than DNN, its high stddev from multiple trials has driven up the mean loss. xDeepFM also suffers from trainability issue (see its high stddev on MovieLens). In xDeepFM, each feature map encodes all the pair-wise crosses while only relies on a single variable to learn the importance of each cross. In practice, a single variable is difficult to be learned when jointly trained with magnitudes more parameters. Then, an improperly learned variable would lead to noises.


{DCN-V2}, on the other hand, consistently outperforms DNN. It successfully leveraged both the explicit and implicit feature interactions. We attribute this to the balanced number of parameters between the cross network and the deep network ({\bf expressive}), as well as the simple structure of cross net which eased the optimization ({\bf easy-to-optimize}). It's worth noting that the high-level structure of {DCN-V2} shares a similar spirit of the self-attention mechanism adopted in AutoInt, where each feature embedding attends to a weighed combination of other features. The difference is that during the attention, higher-order interactions were modeled explicitly in {DCN-V2} but implicitly in AutoInt.

{\bf Model Efficiency.} \autoref{tab:baseline_comparison} also provides details for model size and FLOPS\footnote{FLOPS is a close estimation of run time, which is subjective to implementation details.}. The reported setting was properly tuned over the hyper-parameters of each model and the DNN component. For most models, the FLOPS is roughly 2x of the \#params; for xDeepFM, however, the FLOPS is one magnitude higher, making it impractical in industrial-scale applications (also observed in \cite{song2019autoint}). Note that for DeepFM and DLRM, we've also searched over larger-capacity models; however, they didn't deliver better quality. Among all the methods, {DCN-V2} delivers the best performance while remaining relatively efficient; {DCN-Mix} further reduced the cost, achieving a better trade-off between model efficiency and quality.

\begin{table*}[htpb]
\small
\caption{LogLoss and AUC (test) on Criteo and Movielen-1M. The metrics were averaged over 5 independent runs with their stddev in the parenthesis. In the `Best Setting' column, the left reports DNN setting and the right reports model-specific setting. $l$ denotes layer depth; $n$ denotes CIN layer size; $h$ and $e$, respectively, denotes \#heads and att-embed-size; $K$ denotes \#experts and $r$ denotes total rank.}
\vspace{-3ex}
\label{tab:baseline_comparison}
\begin{center}
\begin{tabular}{l|ccp{2.5em}p{2.5em}ll|ccp{2.5em}p{2.5em}}
\toprule
\multirow{ 2}{*}{{\bf Baseline}} & \multicolumn{6}{c|}{\bf Criteo} & \multicolumn{4}{c}{\bf MovieLens-1M} \\
& Logloss  & AUC & Params & FLOPS &  \multicolumn{2}{c|}{Best Setting} &Logloss & AUC  & Params & FLOPS  \\
\midrule
PNN     & 0.4421 (5.8E-4) & 0.8099 (6.1E-4) & 3.1M & 6.1M & (3, 1024) & OPNN
         & 0.3182 (1.4E-3) & 0.8955 (3.3E-4) & 54K & 110K\\
DeepFm  & 0.4420 (1.4E-4) & 0.8099 (1.5E-4) & 1.4M & 2.8M & (2, 768) & \multicolumn{1}{c|}{--}
         & 0.3202 (1.0E-3) & 0.8932 (7.7E-4) & 46K & 93K\\
DLRM    & 0.4427 (3.1E-4) & 0.8092 (3.1E-4) & 1.1M & 2.2M & (2, 768) &  [512,256,64]
         & 0.3245 (1.1E-3) & 0.8890 (1.1E-3) & 7.7K & 16K\\
xDeepFm & 0.4421 (1.6E-4) & 0.8099 (1.8E-4) & 3.7M & 32M & (3, 1024) & $l$=2, $n$=100
         & 0.3251 (4.3E-3) & 0.8923 (8.6E-4) & 160K & 990K\\
AutoInt+ & 0.4420 (5.7E-5) & 0.8101 (2.6E-5) & 4.2M & 8.7M & (4, 1024) & $l$=2, $h$=2, $e$=40
         & 0.3204 (4.4E-4) & 0.8928 (3.9E-4) & 260K & 500K\\
DCN     & 0.4420 (1.6E-4) & 0.8099 (1.7E-4) & 2.1M & 4.2M & (2, 1024) & $l$=4
         & 0.3197 (1.9E-4) & 0.8935 (2.1E-4) & 110K & 220K\\
DNN    & 0.4421 (6.5E-5) & 0.8098 (5.9E-5) & 3.2M & 6.3M & (3, 1024) &  \multicolumn{1}{c|}{--}
         & 0.3201 (4.1E-4) & 0.8929 (2.3E-4) & 46K & 92K\\
\midrule
\multicolumn{1}{c}{\bf Ours}& & & & & & \multicolumn{1}{c}{}& & & &\\
{DCN-V2}  & \bf 0.4406 (6.2E-5) & \bf 0.8115 (7.1E-5) & 3.5M & 7.0M & (2, 768) & $l$=2
         & 0.3170 (3.6E-4) & 0.8950 (2.7E-4) & 110K & 220K \\
{DCN-Mix} & 0.4408 (1.0E-4) & 0.8112 (9.8E-5) & 2.4M & 4.8M  & (2, 512) & $l$=3, $K$=4, $r$=258
         & \bf 0.3160 (4.9E-4) & \bf 0.8964 (2.9E-4) & 110K & 210K \\
CrossNet &0.4413 (2.5E-4) & 0.8107 (2.4E-4) & 2.1M & 4.2M &  -- & $l$=4, $K$=4, $r$=258
         & 0.3185 (3.0E-4) & 0.8937 (2.7E-4) & 65K & 130K \\
\bottomrule
\end{tabular}
\end{center}
\vspace{-5pt}
\end{table*}

\subsection{Can Cross Layers Replace ReLU layers?}
\label{sec:crossnet_performance}
The solid performance of {DCN-V2} approaches has inspired us to further study the efficiency of their cross layers (CrossNet) in learning explicit high-order feature crosses.

In a realistic setting with resource constraints, we often have to limit model capacity. Hence, we fixed the model capacity (memory / \# of parameters) at different levels, and compared the performance between a model with only cross layers (Cross Net), and a ReLU based DNN. \autoref{tab:logloss_memory} reports the best test LogLoss for different memory constraints. The memory was controlled by varying the number of cross layers and its rank (\{128, 256\}), the number of hidden layers and their sizes. The best performance was achieved by the cross network (5-layer), suggesting the ground-truth model could be well-approximated by polynomials. Moreover, the best performance per memory limit was also achieved by the cross network, indicating both solid effectiveness and efficiency.

It is well known that ReLU layers are the backbone for various Neural Nets models including DNN, Recurrent Neural Net (RNN) \cite{rumelhart1985learning, hochreiter1997long, mikolov2011extensions} and Convolutional Neural Net (CNN) \cite{lecun1989backpropagation, schmidhuber2015deep, lawrence1997face}. It is quite surprising and encouraging to us that we may potentially replace ReLU layers by Cross Layers entirely for certain applications. Obviously we need significant more analysis and experiments to verify the hypothesis. Nonetheless, this is a very interesting preliminary study and sheds light for our future explorations on cross layers.

\begin{table}[htbp]
\small
\vspace{-4pt}
\caption{Logloss and AUC (test) with a fixed memory budget.}
\vspace{-10pt}
\label{tab:logloss_memory}
\begin{center}
\begin{tabular}{cc|cccc}
\toprule
\multicolumn{2}{c|}{\bf \#Params} & 7.9E+05 & 1.3E+06 & 2.1E+06 & 2.6E+06\\
\midrule
\multirow{ 2}{*}{LogLoss}& CrossNet & \bf 0.4424 & \bf 0.4417 & \bf 0.4416 & \bf 0.4415\\
 &  DNN & 0.4427 & 0.4426 & 0.4423 & 0.4423\\
 \midrule
\multirow{ 2}{*}{AUC} &  CrossNet & \bf 0.8096 & \bf 0.8104 & \bf 0.8105  & \bf 0.8106\\
& DNN & 0.8091 & 0.8094  & 0.8096 & 0.80961 \\

\bottomrule
\end{tabular}
\end{center}
\vspace{-4pt}
\end{table}

\subsection{How the Choice of Hyper-parameters Affect {DCN-V2} Model Performance (RQ4)} 
\label{sec:hyper-parameters}
This section examines the model performance as a function of hyper-parameters that include 1) depth of cross layers; 2) matrix rank of {DCN-Mix}; 3) number of experts in {DCN-Mix}.

{\bf Depth of Cross Layers.}
By design, the highest feature cross order captured by the cross net increases with layer depth. Hence, we constrain ourselves to the full-rank cross layers, and evaluate the performance change with layer depth

\begin{figure}[htbp]
\small
\centering
    \begin{subfigure}[b]{0.22\textwidth}  
    \includegraphics[width=\textwidth]{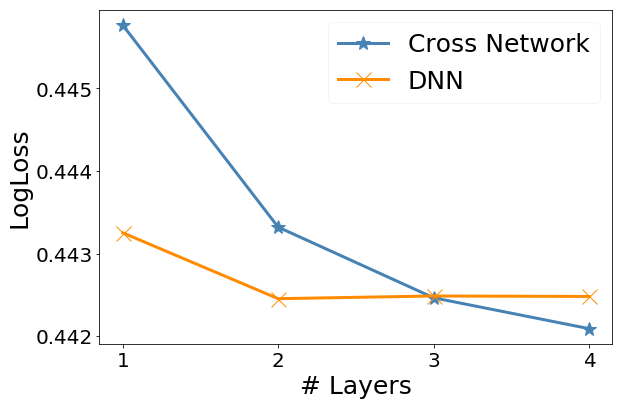}
    \end{subfigure}
    \hfill
    \begin{subfigure}[b]{0.22\textwidth}  
    \includegraphics[width=\textwidth]{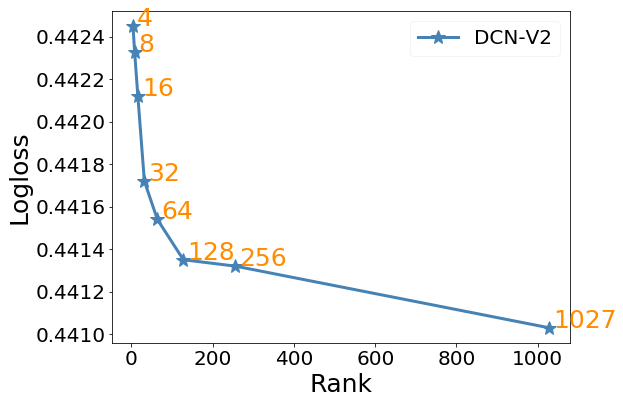}
    \end{subfigure}\\[-2ex]
    \begin{subfigure}[b]{0.22\textwidth}  
    \includegraphics[width=\textwidth]{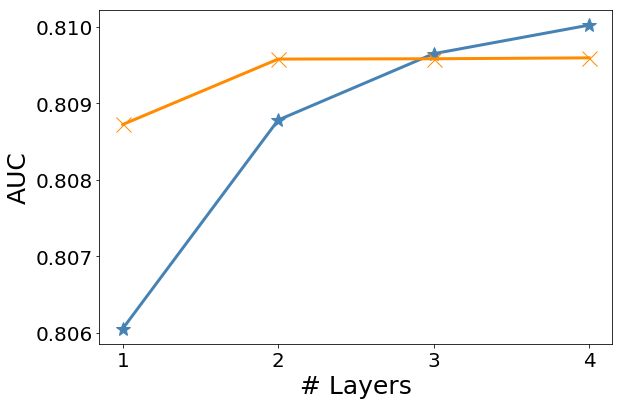}
    \caption{Layer depth}
    \label{fig:dcn-layer-depth}
    \end{subfigure}
    \hfill
    \begin{subfigure}[b]{0.22\textwidth}  
    \includegraphics[width=\textwidth]{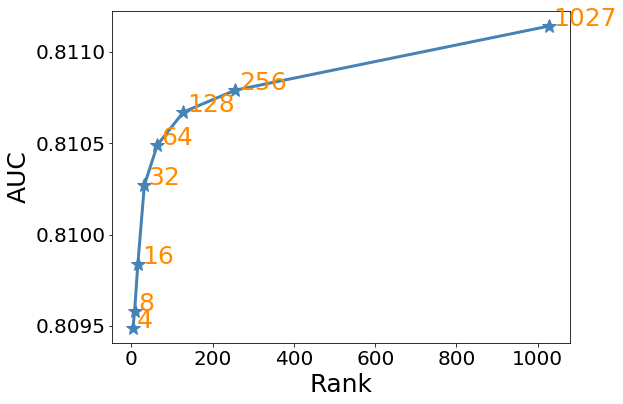}
    \caption{Matrix rank}
    \label{fig:dcn-rank}
    \end{subfigure}
    \vspace{-2ex}
    \caption{Logloss and AUC (test) v.s. depth \& matrix rank.}
\vspace{-10pt}
\end{figure}

\autoref{fig:dcn-layer-depth} shows the test LogLoss and AUC while increasing layer depth on the Criteo dataset. We see a steady quality improvement with a deeper cross network, indicating that it's able to capture more meaningful crosses. The rate of improvement, however, slowed down when more layers were used. This suggests the contribution from that of higher-order crosses is less significant than those from lower-order crosses. We also used a same-sized DNN as a reference. When there were $\le 2$ layers, DNN outperformed the cross network; when more layers became available, the cross network started to close the performance gap and even outperformed DNN. In the small-layer regime, the cross network could only approximate very low-order crosses (\emph{e.g.,} 1 $\sim$ 2); in the large-layer regime, those low-order crosses were characterized with more parameters, and those high-order interactions were started to be captured.

{\bf Rank of Matrix.}
The rank of the weight matrix controls the number of parameters as well as the portion of low-frequency signals passing through the cross layers. Hence, we study its influence on model quality. The model is based on a well-performed setting with 3 cross layers followed by 3 hidden layers of size 512. We approximate the dense matrix $W$ in each cross layer by $UV^\top$ where $U, V \in \mathbb{R}^{d \times r}$, and we vary $r$. We loosely consider the smaller dimension $r$ to be the rank.

\autoref{fig:dcn-rank} shows the test LogLoss and AUC v.s. matrix's rank $r$ on Criteo. When $r$ was as small as 4, the performance was on-par with other baselines. When $r$ was increased from 4 to 64, the LogLoss decreased almost linearly with $r$ (\emph{i.e.}, model's improving). When $r$ was further increased from 64 to full, the improvement on LogLoss slowed down. We refer to 64 as the \emph{threshold rank}. The significant slow down from 64 suggests that the important signals characterizing feature crosses could be captured in the top-64 singular values.

Our hypothesis for the value of this \emph{threshold rank} is $O(k)$ where $k$ represents \# features (39 for Criteo). Consider the $(i,j)$-th block of matrix $W$, we can view $\small W_{i,j} = W_{i,j}^L + W_{i,j}^H$, where $W_{i,j}^L$ stores the dominant signal (low-frequency) and $\small W_{i,j}^H$ stores the rest (high-frequency). In the simplest case where $\small W_{i,j}^L = c_{ij} {\bf 1} {\bf 1}^\top$, the entire matrix $\small W^L$ will be of rank $k$. The effectiveness of this hypothesis remains to be verified across multiple datasets.

{\bf Number of Experts.}
We study how the number of low-rank experts affects the quality. We've observed that 1) best-performed setting (\#expert, gate, matrix activation type) was subjective to datasets and model architectures; 2) the best-performed model of each setting yielded similar results. For example, for a 2-layered cross net with total rank 256 on Criteo, the LogLoss for 1, 4, 8, 16, and 32 experts, respectively, was 0.4418, 0.4416, 0.4416, 0.4422, and 0.4420. The fact that more lower-ranked experts wasn't performing better than a single higher-ranked expert might be caused by the na\"ive gating functions and optimizations adopted. We believe more sophisticated gating \cite{jang2016categorical, louizos2017learning, ma2019snr} and optimization techniques (\emph{e.g.}, alternative training, special initialization, temperature adjustment) would leverage more from a mixture of experts. This, however, is beyond the scope of this paper and we leave it to future work.

\subsection{Model Understanding (RQ5)}
\label{sec:model_understanding}
One key research question is whether the proposed approaches are indeed learning meaningful feature crosses. A good understanding about the learned feature crosses helps improve model understandability, and is especially crucial to fields like ML fairness and ML for health. Fortunately, the weight matrix $W$ in {DCN-V2} exactly reveals what feature crosses the model has learned to be important. Specifically, we assume that each input $\vecx = [\vecx_1; \vecx_2; \ldots; \vecx_k]$ contains $k$ features with each represented by an embedding $\vecx_i$. Then, the block-wise view of the feature crossing component (ignoring the bias) in Eq. \eqref{eq:block-mat} shows that the importance of feature interaction between $i$-th and $j$-th feature is characterized by the $(i,j)$-th block $W_{i,j}$.
\begin{equation} \label{eq:block-mat}
    \vecx \odot W \vecx = 
    \left[\begin{smallmatrix}
        \vecx_1 \\
        \vecx_2 \vspace{-1ex}\\
        \vdots  \\
        \vecx_k
    \end{smallmatrix}\right]
    \odot
    \left[\begin{smallmatrix}
        W_{1,1} & W_{1,2} & \cdots & W_{1,k} \\
        W_{2,1} & W_{2,2} & \cdots & W_{2,k} \vspace{-1ex}\\
        \vdots  & \vdots  & \ddots & \vdots  \\
        W_{k,1} & W_{k,2} & \cdots & W_{k,k}
    \end{smallmatrix}\right]
    \left[\begin{smallmatrix}
        \vecx_1 \\
        \vecx_2 \vspace{-1ex}\\
        \vdots  \\
        \vecx_k
    \end{smallmatrix}\right]
\end{equation}

\begin{figure}[htbp]
\centering
    \begin{subfigure}[b]{0.18\textwidth}  
    \includegraphics[width=\textwidth,valign=t]{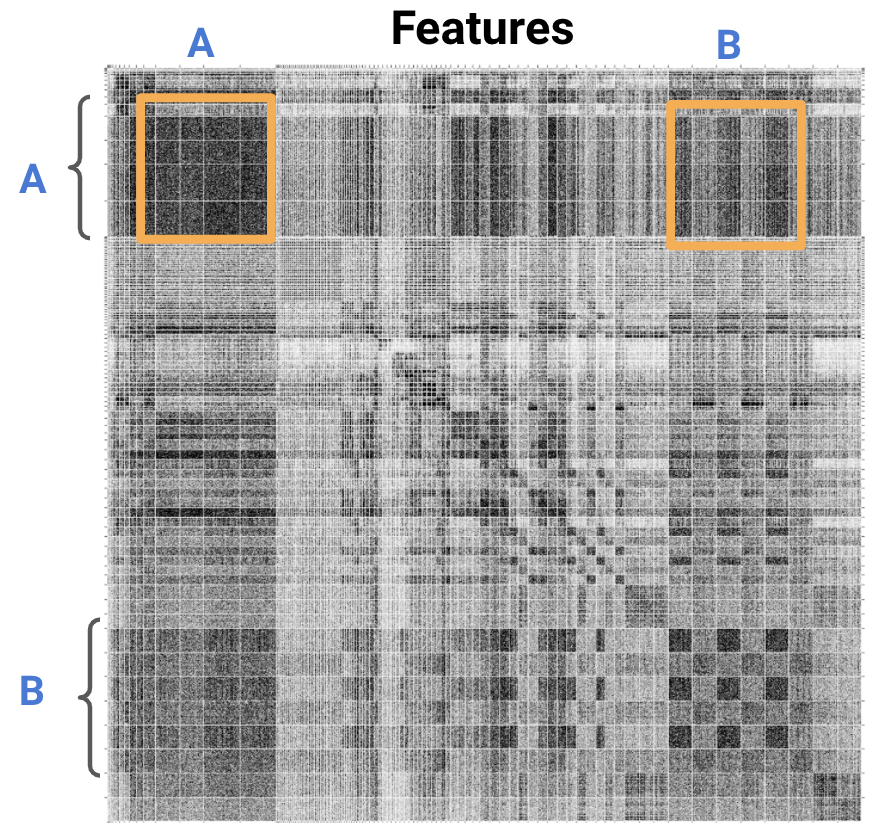}
    \caption{Production data}
    \end{subfigure}
    \hfill
    \begin{subfigure}[b]{0.22\textwidth}  
    \includegraphics[width=\textwidth,valign=t]{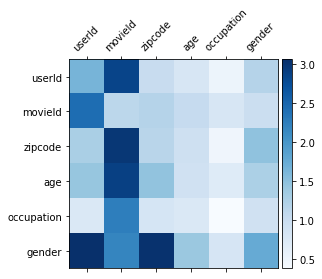}
    \caption{Movielen-1M}
    \end{subfigure}
    \caption{Visualization of learned weight matrix in {DCN-V2}. Rows and columns represents real features. For (a), feature names were not shown for proprietary reasons; darker pixel represents larger weight in its absolute value. For (b), each block represents the Frobenius norm of each matrix block.}
\label{fig:dcn-mat-visualization}
\end{figure}

\autoref{fig:dcn-mat-visualization} shows the learned weight matrix $W$ in the first cross layer. Subplot (a) shows the entire matrix with orange boxes highlighting some notable feature crosses. The off-diagonal block corresponds to crosses that are known to be important, suggesting the effectiveness of {DCN-V2}. The diagonal block represents self-interaction ( $x^2$'s).
Subplot (b) shows each block's Frobenius norm and indicates some strong interactions learned, \emph{e.g.}, {\ttfamily{\small Gender}} $\times$ {\ttfamily{\small UserId}}, {\ttfamily{\small MovieId}} $\times$ {\ttfamily{\small UserId}}.

\section{Productionizing {DCN-V2} at Google}
\label{sec:productionization}
This section provides a case study to share our experience productionizing {DCN-V2} in a large-scale recommender system in Google. We've achieved significant gains through {DCN-V2} in both offline model accuracy, and online key business metrics.

{\bf The Ranking Problem:}
Given a user and a large set of candidates, our problem is to return the top-$k$ items the user is most likely to engage with. Let's denote the training data to be $\{(\vecx_i, y_i)\}_{i=1}^N$, where $\vecx_i$'s represents features of multiple modalities, such as user's interests, an item's metadata and contextual features; $y_i$'s are labels representing a user's action (\emph{e.g.}, a click). The goal is to learn a function $f: \mathbb{R}^d \mapsto \mathbb{R}$ that predicts the probability $P(y | \vecx)$, the user's action $y$ given features $\vecx$.

{\bf Production Data and Model:}
The production data are sampled user logs consisting of hundreds of billions of training examples. The vocabulary sizes of sparse features vary from 2 to millions. The baseline model is a fully-connected multi-layer perceptron (MLP) with ReLU activations.




{\bf Comparisons with Production Models:} When compared with production model, {DCN-V2} yielded 0.6\% AUCLoss (1 - AUC) improvement. For this particular model, a gain of 0.1\% on AUCLoss is considered a significant improvement.  We also observed significant online performance gains on key metrics.
\autoref{tab:dcn-relu-prod} further verifies the amount of gain from {DCN-V2} by replacing cross layers with same-sized ReLU layers.

\begin{table}[htpb]
\small
    \caption{Relative AUCLoss of {DCN-V2} v.s. same-sized ReLUs}
    \label{tab:dcn-relu-prod}
    \vspace{-3ex}
        \centering
        \begin{tabular}{cccc}
        \toprule
        { 1layer ReLU}  & {2layer ReLU} & {1layer {DCN-V2}} & {2layer {DCN-V2}}\\
        \midrule
         0\%   & -0.15\% & -0.19\% &  -0.45\% \\
        \bottomrule
        \end{tabular}
\end{table}

{\bf Practical Learnings.} We share some practical lessons we have learned through productionizing {DCN-V2}. 
\begin{itemize}[leftmargin=1em]
\item It's better to insert the cross layers in between the input and the hidden layers of DNN (also observed in \cite{shan2016deep}). Our hypothesis is that the physical meaning of feature representations and their interactions becomes weaker as it goes farther away from the input layer.
\item We saw consistent accuracy gains by stacking or concatenating 1 - 2 cross layers. Beyond 2 cross layers, the gains start to plateau.
\item We observed that both stacking cross layers and concatenating cross layers work well. Stacking layers learns higher-order feature interactions, while concatenating layers (similar to multi-head mechanism \cite{vaswani2017attention}) captures complimentary interactions.
\item We observed that using low-rank DCN with rank $\text{(input size)}/4$ consistently preserved the accuracy of a full-rank {DCN-V2}.  
\end{itemize}

\section{Conclusions and Future Work}

In this paper, we propose a new model---{DCN-V2}---to model explicit crosses in an expressive yet simple manner. Observing the low-rank nature of the weight matrix in the cross network, we also propose a mixture of low-rank DCN ({DCN-Mix}) to achieve a healthier trade-off between model performance and latency. {DCN-V2} has been successfully deployed in multiple web-scale learning to rank systems with significant offline model accuracy and online business metric gains. Our experimental results also have demonstrated {DCN-V2}'s effectiveness over SOTA methods.


For future work, we are interested in advancing our understanding of 1). the interactions between {DCN-V2} and optimization algorithms such as second-order methods; 2). the relation between embedding, {DCN-V2} and its rank of matrix. Further, we would like to improve the gating mechanism in {DCN-Mix}. Moreover, observing that cross layers in {DCN-V2} may serve as potential alternatives to ReLU layers in DNNs, we are very interested to verify this observation across more complex model architectures (\emph{e.g.}, RNN, CNN).

\medskip
{\bf Acknowledgement.} We would like to thank Bin Fu, Gang (Thomas) Fu, and Mingliang Wang for their early contributions of {DCN-V2}; Tianshuo Deng, Wenjing Ma, Yayang Tian, Shuying Zhang, Jie (Jerry) Zhang, Evan Ettinger, Samuel Ieong and many others for their efforts and supports in productionizing {DCN-V2}; Ting Chen for his initial idea of mixture of low-rank; and Jiaxi Tang for his valuable comments.

\bibliographystyle{ACM-Reference-Format}
\bibliography{main}


\begin{thebibliography}{00}


\ifx \showCODEN    \undefined \def \showCODEN     #1{\unskip}     \fi
\ifx \showDOI      \undefined \def \showDOI       #1{#1}\fi
\ifx \showISBNx    \undefined \def \showISBNx     #1{\unskip}     \fi
\ifx \showISBNxiii \undefined \def \showISBNxiii  #1{\unskip}     \fi
\ifx \showISSN     \undefined \def \showISSN      #1{\unskip}     \fi
\ifx \showLCCN     \undefined \def \showLCCN      #1{\unskip}     \fi
\ifx \shownote     \undefined \def \shownote      #1{#1}          \fi
\ifx \showarticletitle \undefined \def \showarticletitle #1{#1}   \fi
\ifx \showURL      \undefined \def \showURL       {\relax}        \fi
\providecommand\bibfield[2]{#2}
\providecommand\bibinfo[2]{#2}
\providecommand\natexlab[1]{#1}
\providecommand\showeprint[2][]{arXiv:#2}

\bibitem[\protect\citeauthoryear{Beutel, Covington, Jain, Xu, Li, Gatto, and
  Chi}{Beutel et~al\mbox{.}}{2018}]%
        {beutel2018latent}
\bibfield{author}{\bibinfo{person}{Alex Beutel}, \bibinfo{person}{Paul
  Covington}, \bibinfo{person}{Sagar Jain}, \bibinfo{person}{Can Xu},
  \bibinfo{person}{Jia Li}, \bibinfo{person}{Vince Gatto}, {and}
  \bibinfo{person}{Ed~H Chi}.} \bibinfo{year}{2018}\natexlab{}.
\newblock \showarticletitle{Latent cross: Making use of context in recurrent
  recommender systems}. In \bibinfo{booktitle}{{\em Proceedings of the Eleventh
  ACM International Conference on Web Search and Data Mining}}.
  \bibinfo{pages}{46--54}.
\newblock


\bibitem[\protect\citeauthoryear{Bottou, Peters, Qui{\~n}onero-Candela,
  Charles, Chickering, Portugaly, Ray, Simard, and Snelson}{Bottou
  et~al\mbox{.}}{2013}]%
        {bottou2013counterfactual}
\bibfield{author}{\bibinfo{person}{L{\'e}on Bottou}, \bibinfo{person}{Jonas
  Peters}, \bibinfo{person}{Joaquin Qui{\~n}onero-Candela},
  \bibinfo{person}{Denis~X Charles}, \bibinfo{person}{D~Max Chickering},
  \bibinfo{person}{Elon Portugaly}, \bibinfo{person}{Dipankar Ray},
  \bibinfo{person}{Patrice Simard}, {and} \bibinfo{person}{Ed Snelson}.}
  \bibinfo{year}{2013}\natexlab{}.
\newblock \showarticletitle{Counterfactual reasoning and learning systems: The
  example of computational advertising}.
\newblock \bibinfo{journal}{{\em The Journal of Machine Learning Research\/}}
  \bibinfo{volume}{14}, \bibinfo{number}{1} (\bibinfo{year}{2013}),
  \bibinfo{pages}{3207--3260}.
\newblock


\bibitem[\protect\citeauthoryear{Broder}{Broder}{2008}]%
        {broder2008computational}
\bibfield{author}{\bibinfo{person}{Andrei~Z Broder}.}
  \bibinfo{year}{2008}\natexlab{}.
\newblock \showarticletitle{Computational advertising and recommender systems}.
  In \bibinfo{booktitle}{{\em Proceedings of the 2008 ACM conference on
  Recommender systems}}. \bibinfo{pages}{1--2}.
\newblock


\bibitem[\protect\citeauthoryear{Cao, Qin, Liu, Tsai, and Li}{Cao
  et~al\mbox{.}}{2007}]%
        {cao2007learning}
\bibfield{author}{\bibinfo{person}{Zhe Cao}, \bibinfo{person}{Tao Qin},
  \bibinfo{person}{Tie-Yan Liu}, \bibinfo{person}{Ming-Feng Tsai}, {and}
  \bibinfo{person}{Hang Li}.} \bibinfo{year}{2007}\natexlab{}.
\newblock \showarticletitle{Learning to rank: from pairwise approach to
  listwise approach}. In \bibinfo{booktitle}{{\em Proceedings of the 24th
  international conference on Machine learning}}. \bibinfo{pages}{129--136}.
\newblock


\bibitem[\protect\citeauthoryear{Chen, Lin, Lin, Han, Wang, and Zhou}{Chen
  et~al\mbox{.}}{2018}]%
        {chen2018adaptive}
\bibfield{author}{\bibinfo{person}{Ting Chen}, \bibinfo{person}{Ji Lin},
  \bibinfo{person}{Tian Lin}, \bibinfo{person}{Song Han},
  \bibinfo{person}{Chong Wang}, {and} \bibinfo{person}{Denny Zhou}.}
  \bibinfo{year}{2018}\natexlab{}.
\newblock \showarticletitle{Adaptive mixture of low-rank factorizations for
  compact neural modeling}.
\newblock  (\bibinfo{year}{2018}).
\newblock


\bibitem[\protect\citeauthoryear{Cheng, Koc, Harmsen, Shaked, Chandra, Aradhye,
  Anderson, Corrado, Chai, Ispir, et~al\mbox{.}}{Cheng et~al\mbox{.}}{2016}]%
        {cheng2016wide}
\bibfield{author}{\bibinfo{person}{Heng-Tze Cheng}, \bibinfo{person}{Levent
  Koc}, \bibinfo{person}{Jeremiah Harmsen}, \bibinfo{person}{Tal Shaked},
  \bibinfo{person}{Tushar Chandra}, \bibinfo{person}{Hrishi Aradhye},
  \bibinfo{person}{Glen Anderson}, \bibinfo{person}{Greg Corrado},
  \bibinfo{person}{Wei Chai}, \bibinfo{person}{Mustafa Ispir}, {et~al\mbox{.}}}
  \bibinfo{year}{2016}\natexlab{}.
\newblock \showarticletitle{Wide \& Deep Learning for Recommender Systems}.
\newblock \bibinfo{journal}{{\em arXiv preprint arXiv:1606.07792\/}}
  (\bibinfo{year}{2016}).
\newblock


\bibitem[\protect\citeauthoryear{Cheng, Shen, and Huang}{Cheng
  et~al\mbox{.}}{2019}]%
        {cheng2019adaptive}
\bibfield{author}{\bibinfo{person}{Weiyu Cheng}, \bibinfo{person}{Yanyan Shen},
  {and} \bibinfo{person}{Linpeng Huang}.} \bibinfo{year}{2019}\natexlab{}.
\newblock \showarticletitle{Adaptive Factorization Network: Learning
  Adaptive-Order Feature Interactions}.
\newblock \bibinfo{journal}{{\em arXiv preprint arXiv:1909.03276\/}}
  (\bibinfo{year}{2019}).
\newblock


\bibitem[\protect\citeauthoryear{Dacrema, Cremonesi, and Jannach}{Dacrema
  et~al\mbox{.}}{2019}]%
        {dacrema2019we}
\bibfield{author}{\bibinfo{person}{Maurizio~Ferrari Dacrema},
  \bibinfo{person}{Paolo Cremonesi}, {and} \bibinfo{person}{Dietmar Jannach}.}
  \bibinfo{year}{2019}\natexlab{}.
\newblock \showarticletitle{Are we really making much progress? A worrying
  analysis of recent neural recommendation approaches}. In
  \bibinfo{booktitle}{{\em Proceedings of the 13th ACM Conference on
  Recommender Systems}}. \bibinfo{pages}{101--109}.
\newblock


\bibitem[\protect\citeauthoryear{Drineas and Mahoney}{Drineas and
  Mahoney}{2005}]%
        {drineas2005nystrom}
\bibfield{author}{\bibinfo{person}{Petros Drineas} {and}
  \bibinfo{person}{Michael~W Mahoney}.} \bibinfo{year}{2005}\natexlab{}.
\newblock \showarticletitle{On the Nystr{\"o}m method for approximating a Gram
  matrix for improved kernel-based learning}.
\newblock \bibinfo{journal}{{\em journal of machine learning research\/}}
  \bibinfo{volume}{6}, \bibinfo{number}{Dec} (\bibinfo{year}{2005}),
  \bibinfo{pages}{2153--2175}.
\newblock


\bibitem[\protect\citeauthoryear{Eigen, Ranzato, and Sutskever}{Eigen
  et~al\mbox{.}}{2013}]%
        {eigen2013learning}
\bibfield{author}{\bibinfo{person}{David Eigen}, \bibinfo{person}{Marc'Aurelio
  Ranzato}, {and} \bibinfo{person}{Ilya Sutskever}.}
  \bibinfo{year}{2013}\natexlab{}.
\newblock \showarticletitle{Learning factored representations in a deep mixture
  of experts}.
\newblock \bibinfo{journal}{{\em arXiv preprint arXiv:1312.4314\/}}
  (\bibinfo{year}{2013}).
\newblock


\bibitem[\protect\citeauthoryear{Fan, Feliu-Faba, Lin, Ying, and
  Zepeda-N{\'u}nez}{Fan et~al\mbox{.}}{2019}]%
        {fan2019multiscale}
\bibfield{author}{\bibinfo{person}{Yuwei Fan}, \bibinfo{person}{Jordi
  Feliu-Faba}, \bibinfo{person}{Lin Lin}, \bibinfo{person}{Lexing Ying}, {and}
  \bibinfo{person}{Leonardo Zepeda-N{\'u}nez}.}
  \bibinfo{year}{2019}\natexlab{}.
\newblock \showarticletitle{A multiscale neural network based on hierarchical
  nested bases}.
\newblock \bibinfo{journal}{{\em Research in the Mathematical Sciences\/}}
  \bibinfo{volume}{6}, \bibinfo{number}{2} (\bibinfo{year}{2019}),
  \bibinfo{pages}{21}.
\newblock


\bibitem[\protect\citeauthoryear{Golub and Van~Loan}{Golub and
  Van~Loan}{1996}]%
        {golub1996matrix}
\bibfield{author}{\bibinfo{person}{Gene~H Golub} {and}
  \bibinfo{person}{Charles~F Van~Loan}.} \bibinfo{year}{1996}\natexlab{}.
\newblock \showarticletitle{Matrix Computations Johns Hopkins University
  Press}.
\newblock \bibinfo{journal}{{\em Baltimore and London\/}}
  (\bibinfo{year}{1996}).
\newblock


\bibitem[\protect\citeauthoryear{Guo, Tang, Ye, Li, and He}{Guo
  et~al\mbox{.}}{2017}]%
        {guo2017deepfm}
\bibfield{author}{\bibinfo{person}{Huifeng Guo}, \bibinfo{person}{Ruiming
  Tang}, \bibinfo{person}{Yunming Ye}, \bibinfo{person}{Zhenguo Li}, {and}
  \bibinfo{person}{Xiuqiang He}.} \bibinfo{year}{2017}\natexlab{}.
\newblock \showarticletitle{DeepFM: a factorization-machine based neural
  network for CTR prediction}.
\newblock \bibinfo{journal}{{\em arXiv preprint arXiv:1703.04247\/}}
  (\bibinfo{year}{2017}).
\newblock


\bibitem[\protect\citeauthoryear{Halko, Martinsson, and Tropp}{Halko
  et~al\mbox{.}}{2011}]%
        {halko2011finding}
\bibfield{author}{\bibinfo{person}{Nathan Halko}, \bibinfo{person}{Per-Gunnar
  Martinsson}, {and} \bibinfo{person}{Joel~A Tropp}.}
  \bibinfo{year}{2011}\natexlab{}.
\newblock \showarticletitle{Finding structure with randomness: Probabilistic
  algorithms for constructing approximate matrix decompositions}.
\newblock \bibinfo{journal}{{\em SIAM review\/}} \bibinfo{volume}{53},
  \bibinfo{number}{2} (\bibinfo{year}{2011}), \bibinfo{pages}{217--288}.
\newblock


\bibitem[\protect\citeauthoryear{He, Zhang, Ren, and Sun}{He
  et~al\mbox{.}}{2015}]%
        {he2015delving}
\bibfield{author}{\bibinfo{person}{Kaiming He}, \bibinfo{person}{Xiangyu
  Zhang}, \bibinfo{person}{Shaoqing Ren}, {and} \bibinfo{person}{Jian Sun}.}
  \bibinfo{year}{2015}\natexlab{}.
\newblock \showarticletitle{Delving deep into rectifiers: Surpassing
  human-level performance on imagenet classification}. In
  \bibinfo{booktitle}{{\em Proceedings of the IEEE international conference on
  computer vision}}. \bibinfo{pages}{1026--1034}.
\newblock


\bibitem[\protect\citeauthoryear{He and Chua}{He and Chua}{2017}]%
        {he2017neural}
\bibfield{author}{\bibinfo{person}{Xiangnan He} {and} \bibinfo{person}{Tat-Seng
  Chua}.} \bibinfo{year}{2017}\natexlab{}.
\newblock \showarticletitle{Neural factorization machines for sparse predictive
  analytics}. In \bibinfo{booktitle}{{\em Proceedings of the 40th International
  ACM SIGIR conference on Research and Development in Information Retrieval}}.
  \bibinfo{pages}{355--364}.
\newblock


\bibitem[\protect\citeauthoryear{Herlocker, Konstan, Terveen, and
  Riedl}{Herlocker et~al\mbox{.}}{2004}]%
        {herlocker2004evaluating}
\bibfield{author}{\bibinfo{person}{Jonathan~L Herlocker},
  \bibinfo{person}{Joseph~A Konstan}, \bibinfo{person}{Loren~G Terveen}, {and}
  \bibinfo{person}{John~T Riedl}.} \bibinfo{year}{2004}\natexlab{}.
\newblock \showarticletitle{Evaluating collaborative filtering recommender
  systems}.
\newblock \bibinfo{journal}{{\em ACM Transactions on Information Systems
  (TOIS)\/}} \bibinfo{volume}{22}, \bibinfo{number}{1} (\bibinfo{year}{2004}),
  \bibinfo{pages}{5--53}.
\newblock


\bibitem[\protect\citeauthoryear{Hochreiter and Schmidhuber}{Hochreiter and
  Schmidhuber}{1997}]%
        {hochreiter1997long}
\bibfield{author}{\bibinfo{person}{Sepp Hochreiter} {and}
  \bibinfo{person}{J{\"u}rgen Schmidhuber}.} \bibinfo{year}{1997}\natexlab{}.
\newblock \showarticletitle{Long short-term memory}.
\newblock \bibinfo{journal}{{\em Neural computation\/}} \bibinfo{volume}{9},
  \bibinfo{number}{8} (\bibinfo{year}{1997}), \bibinfo{pages}{1735--1780}.
\newblock


\bibitem[\protect\citeauthoryear{Jacobs, Jordan, Nowlan, and Hinton}{Jacobs
  et~al\mbox{.}}{1991}]%
        {jacobs1991adaptive}
\bibfield{author}{\bibinfo{person}{Robert~A Jacobs}, \bibinfo{person}{Michael~I
  Jordan}, \bibinfo{person}{Steven~J Nowlan}, {and} \bibinfo{person}{Geoffrey~E
  Hinton}.} \bibinfo{year}{1991}\natexlab{}.
\newblock \showarticletitle{Adaptive mixtures of local experts}.
\newblock \bibinfo{journal}{{\em Neural computation\/}} \bibinfo{volume}{3},
  \bibinfo{number}{1} (\bibinfo{year}{1991}), \bibinfo{pages}{79--87}.
\newblock


\bibitem[\protect\citeauthoryear{Jaderberg, Vedaldi, and Zisserman}{Jaderberg
  et~al\mbox{.}}{2014}]%
        {jaderberg2014speeding}
\bibfield{author}{\bibinfo{person}{Max Jaderberg}, \bibinfo{person}{Andrea
  Vedaldi}, {and} \bibinfo{person}{Andrew Zisserman}.}
  \bibinfo{year}{2014}\natexlab{}.
\newblock \showarticletitle{Speeding up convolutional neural networks with low
  rank expansions}.
\newblock \bibinfo{journal}{{\em arXiv preprint arXiv:1405.3866\/}}
  (\bibinfo{year}{2014}).
\newblock


\bibitem[\protect\citeauthoryear{Jang, Gu, and Poole}{Jang
  et~al\mbox{.}}{2016}]%
        {jang2016categorical}
\bibfield{author}{\bibinfo{person}{Eric Jang}, \bibinfo{person}{Shixiang Gu},
  {and} \bibinfo{person}{Ben Poole}.} \bibinfo{year}{2016}\natexlab{}.
\newblock \showarticletitle{Categorical reparameterization with
  gumbel-softmax}.
\newblock \bibinfo{journal}{{\em arXiv preprint arXiv:1611.01144\/}}
  (\bibinfo{year}{2016}).
\newblock


\bibitem[\protect\citeauthoryear{Kingma and Ba}{Kingma and Ba}{2014}]%
        {kingma2014adam}
\bibfield{author}{\bibinfo{person}{Diederik Kingma} {and}
  \bibinfo{person}{Jimmy Ba}.} \bibinfo{year}{2014}\natexlab{}.
\newblock \showarticletitle{Adam: A method for stochastic optimization}.
\newblock \bibinfo{journal}{{\em arXiv preprint arXiv:1412.6980\/}}
  (\bibinfo{year}{2014}).
\newblock


\bibitem[\protect\citeauthoryear{Lawrence, Giles, Tsoi, and Back}{Lawrence
  et~al\mbox{.}}{1997}]%
        {lawrence1997face}
\bibfield{author}{\bibinfo{person}{Steve Lawrence}, \bibinfo{person}{C~Lee
  Giles}, \bibinfo{person}{Ah~Chung Tsoi}, {and} \bibinfo{person}{Andrew~D
  Back}.} \bibinfo{year}{1997}\natexlab{}.
\newblock \showarticletitle{Face recognition: A convolutional neural-network
  approach}.
\newblock \bibinfo{journal}{{\em IEEE transactions on neural networks\/}}
  \bibinfo{volume}{8}, \bibinfo{number}{1} (\bibinfo{year}{1997}),
  \bibinfo{pages}{98--113}.
\newblock


\bibitem[\protect\citeauthoryear{LeCun, Boser, Denker, Henderson, Howard,
  Hubbard, and Jackel}{LeCun et~al\mbox{.}}{1989}]%
        {lecun1989backpropagation}
\bibfield{author}{\bibinfo{person}{Yann LeCun}, \bibinfo{person}{Bernhard
  Boser}, \bibinfo{person}{John~S Denker}, \bibinfo{person}{Donnie Henderson},
  \bibinfo{person}{Richard~E Howard}, \bibinfo{person}{Wayne Hubbard}, {and}
  \bibinfo{person}{Lawrence~D Jackel}.} \bibinfo{year}{1989}\natexlab{}.
\newblock \showarticletitle{Backpropagation applied to handwritten zip code
  recognition}.
\newblock \bibinfo{journal}{{\em Neural computation\/}} \bibinfo{volume}{1},
  \bibinfo{number}{4} (\bibinfo{year}{1989}), \bibinfo{pages}{541--551}.
\newblock


\bibitem[\protect\citeauthoryear{Li, Cheng, Chen, Chen, and Wang}{Li
  et~al\mbox{.}}{2020}]%
        {li2020interpretable}
\bibfield{author}{\bibinfo{person}{Zeyu Li}, \bibinfo{person}{Wei Cheng},
  \bibinfo{person}{Yang Chen}, \bibinfo{person}{Haifeng Chen}, {and}
  \bibinfo{person}{Wei Wang}.} \bibinfo{year}{2020}\natexlab{}.
\newblock \showarticletitle{Interpretable Click-Through Rate Prediction through
  Hierarchical Attention}. In \bibinfo{booktitle}{{\em Proceedings of the 13th
  International Conference on Web Search and Data Mining}}.
  \bibinfo{pages}{313--321}.
\newblock


\bibitem[\protect\citeauthoryear{Lian, Zhou, Zhang, Chen, Xie, and Sun}{Lian
  et~al\mbox{.}}{2018}]%
        {lian2018xdeepfm}
\bibfield{author}{\bibinfo{person}{Jianxun Lian}, \bibinfo{person}{Xiaohuan
  Zhou}, \bibinfo{person}{Fuzheng Zhang}, \bibinfo{person}{Zhongxia Chen},
  \bibinfo{person}{Xing Xie}, {and} \bibinfo{person}{Guangzhong Sun}.}
  \bibinfo{year}{2018}\natexlab{}.
\newblock \showarticletitle{xdeepfm: Combining explicit and implicit feature
  interactions for recommender systems}. In \bibinfo{booktitle}{{\em
  Proceedings of the 24th ACM SIGKDD International Conference on Knowledge
  Discovery \& Data Mining}}. \bibinfo{pages}{1754--1763}.
\newblock


\bibitem[\protect\citeauthoryear{Liu}{Liu}{2011}]%
        {liu2011learning}
\bibfield{author}{\bibinfo{person}{Tie-Yan Liu}.}
  \bibinfo{year}{2011}\natexlab{}.
\newblock \bibinfo{booktitle}{{\em Learning to rank for information
  retrieval}}.
\newblock \bibinfo{publisher}{Springer Science \& Business Media}.
\newblock


\bibitem[\protect\citeauthoryear{Louizos, Welling, and Kingma}{Louizos
  et~al\mbox{.}}{2017}]%
        {louizos2017learning}
\bibfield{author}{\bibinfo{person}{Christos Louizos}, \bibinfo{person}{Max
  Welling}, {and} \bibinfo{person}{Diederik~P Kingma}.}
  \bibinfo{year}{2017}\natexlab{}.
\newblock \showarticletitle{Learning Sparse Neural Networks through $ L\_0 $
  Regularization}.
\newblock \bibinfo{journal}{{\em arXiv preprint arXiv:1712.01312\/}}
  (\bibinfo{year}{2017}).
\newblock


\bibitem[\protect\citeauthoryear{Ma, Zhao, Chen, Li, Hong, and Chi}{Ma
  et~al\mbox{.}}{2019}]%
        {ma2019snr}
\bibfield{author}{\bibinfo{person}{Jiaqi Ma}, \bibinfo{person}{Zhe Zhao},
  \bibinfo{person}{Jilin Chen}, \bibinfo{person}{Ang Li},
  \bibinfo{person}{Lichan Hong}, {and} \bibinfo{person}{Ed~H Chi}.}
  \bibinfo{year}{2019}\natexlab{}.
\newblock \showarticletitle{Snr: Sub-network routing for flexible parameter
  sharing in multi-task learning}. In \bibinfo{booktitle}{{\em Proceedings of
  the AAAI Conference on Artificial Intelligence}}, Vol.~\bibinfo{volume}{33}.
  \bibinfo{pages}{216--223}.
\newblock


\bibitem[\protect\citeauthoryear{Ma, Zhao, Yi, Chen, Hong, and Chi}{Ma
  et~al\mbox{.}}{2018}]%
        {ma2018modeling}
\bibfield{author}{\bibinfo{person}{Jiaqi Ma}, \bibinfo{person}{Zhe Zhao},
  \bibinfo{person}{Xinyang Yi}, \bibinfo{person}{Jilin Chen},
  \bibinfo{person}{Lichan Hong}, {and} \bibinfo{person}{Ed~H Chi}.}
  \bibinfo{year}{2018}\natexlab{}.
\newblock \showarticletitle{Modeling task relationships in multi-task learning
  with multi-gate mixture-of-experts}. In \bibinfo{booktitle}{{\em Proceedings
  of the 24th ACM SIGKDD International Conference on Knowledge Discovery \&
  Data Mining}}. \bibinfo{pages}{1930--1939}.
\newblock


\bibitem[\protect\citeauthoryear{Mhaskar}{Mhaskar}{1996}]%
        {mhaskar1996neural}
\bibfield{author}{\bibinfo{person}{Hrushikesh~N Mhaskar}.}
  \bibinfo{year}{1996}\natexlab{}.
\newblock \showarticletitle{Neural networks for optimal approximation of smooth
  and analytic functions}.
\newblock \bibinfo{journal}{{\em Neural computation\/}} \bibinfo{volume}{8},
  \bibinfo{number}{1} (\bibinfo{year}{1996}), \bibinfo{pages}{164--177}.
\newblock


\bibitem[\protect\citeauthoryear{Mikolov, Kombrink, Burget, {\v{C}}ernock{\`y},
  and Khudanpur}{Mikolov et~al\mbox{.}}{2011}]%
        {mikolov2011extensions}
\bibfield{author}{\bibinfo{person}{Tom{\'a}{\v{s}} Mikolov},
  \bibinfo{person}{Stefan Kombrink}, \bibinfo{person}{Luk{\'a}{\v{s}} Burget},
  \bibinfo{person}{Jan {\v{C}}ernock{\`y}}, {and} \bibinfo{person}{Sanjeev
  Khudanpur}.} \bibinfo{year}{2011}\natexlab{}.
\newblock \showarticletitle{Extensions of recurrent neural network language
  model}. In \bibinfo{booktitle}{{\em 2011 IEEE international conference on
  acoustics, speech and signal processing (ICASSP)}}. IEEE,
  \bibinfo{pages}{5528--5531}.
\newblock


\bibitem[\protect\citeauthoryear{Musgrave, Belongie, and Lim}{Musgrave
  et~al\mbox{.}}{2020}]%
        {musgrave2020metric}
\bibfield{author}{\bibinfo{person}{Kevin Musgrave}, \bibinfo{person}{Serge
  Belongie}, {and} \bibinfo{person}{Ser-Nam Lim}.}
  \bibinfo{year}{2020}\natexlab{}.
\newblock \showarticletitle{A metric learning reality check}.
\newblock \bibinfo{journal}{{\em arXiv preprint arXiv:2003.08505\/}}
  (\bibinfo{year}{2020}).
\newblock


\bibitem[\protect\citeauthoryear{Naumov, Mudigere, Shi, Huang, Sundaraman,
  Park, Wang, Gupta, Wu, Azzolini, et~al\mbox{.}}{Naumov et~al\mbox{.}}{2019}]%
        {naumov2019deep}
\bibfield{author}{\bibinfo{person}{Maxim Naumov}, \bibinfo{person}{Dheevatsa
  Mudigere}, \bibinfo{person}{Hao-Jun~Michael Shi}, \bibinfo{person}{Jianyu
  Huang}, \bibinfo{person}{Narayanan Sundaraman}, \bibinfo{person}{Jongsoo
  Park}, \bibinfo{person}{Xiaodong Wang}, \bibinfo{person}{Udit Gupta},
  \bibinfo{person}{Carole-Jean Wu}, \bibinfo{person}{Alisson~G Azzolini},
  {et~al\mbox{.}}} \bibinfo{year}{2019}\natexlab{}.
\newblock \showarticletitle{Deep learning recommendation model for
  personalization and recommendation systems}.
\newblock \bibinfo{journal}{{\em arXiv preprint arXiv:1906.00091\/}}
  (\bibinfo{year}{2019}).
\newblock


\bibitem[\protect\citeauthoryear{Qu, Cai, Ren, Zhang, Yu, Wen, and Wang}{Qu
  et~al\mbox{.}}{2016}]%
        {qu2016product}
\bibfield{author}{\bibinfo{person}{Yanru Qu}, \bibinfo{person}{Han Cai},
  \bibinfo{person}{Kan Ren}, \bibinfo{person}{Weinan Zhang},
  \bibinfo{person}{Yong Yu}, \bibinfo{person}{Ying Wen}, {and}
  \bibinfo{person}{Jun Wang}.} \bibinfo{year}{2016}\natexlab{}.
\newblock \showarticletitle{Product-based neural networks for user response
  prediction}. In \bibinfo{booktitle}{{\em 2016 IEEE 16th International
  Conference on Data Mining (ICDM)}}. IEEE, \bibinfo{pages}{1149--1154}.
\newblock


\bibitem[\protect\citeauthoryear{Rendle}{Rendle}{2010}]%
        {rendle2010factorization}
\bibfield{author}{\bibinfo{person}{Steffen Rendle}.}
  \bibinfo{year}{2010}\natexlab{}.
\newblock \showarticletitle{Factorization machines}. In
  \bibinfo{booktitle}{{\em 2010 IEEE International Conference on Data Mining}}.
  IEEE, \bibinfo{pages}{995--1000}.
\newblock


\bibitem[\protect\citeauthoryear{Rendle}{Rendle}{2012}]%
        {rendle:tist2012}
\bibfield{author}{\bibinfo{person}{Steffen Rendle}.}
  \bibinfo{year}{2012}\natexlab{}.
\newblock \showarticletitle{Factorization Machines with {libFM}}.
\newblock \bibinfo{journal}{{\em ACM Trans. Intell. Syst. Technol.\/}}
  \bibinfo{volume}{3}, \bibinfo{number}{3}, Article \bibinfo{articleno}{57}
  (\bibinfo{date}{May} \bibinfo{year}{2012}), \bibinfo{numpages}{22}~pages.
\newblock
\showISSN{2157-6904}


\bibitem[\protect\citeauthoryear{Rendle, Krichene, Zhang, and Anderson}{Rendle
  et~al\mbox{.}}{2020}]%
        {rendle2020neural}
\bibfield{author}{\bibinfo{person}{Steffen Rendle}, \bibinfo{person}{Walid
  Krichene}, \bibinfo{person}{Li Zhang}, {and} \bibinfo{person}{John
  Anderson}.} \bibinfo{year}{2020}\natexlab{}.
\newblock \showarticletitle{Neural Collaborative Filtering vs. Matrix
  Factorization Revisited}.
\newblock \bibinfo{journal}{{\em arXiv preprint arXiv:2005.09683\/}}
  (\bibinfo{year}{2020}).
\newblock


\bibitem[\protect\citeauthoryear{Resnick and Varian}{Resnick and
  Varian}{1997}]%
        {resnick1997recommender}
\bibfield{author}{\bibinfo{person}{Paul Resnick} {and} \bibinfo{person}{Hal~R
  Varian}.} \bibinfo{year}{1997}\natexlab{}.
\newblock \showarticletitle{Recommender systems}.
\newblock \bibinfo{journal}{{\it Commun. ACM}} \bibinfo{volume}{40},
  \bibinfo{number}{3} (\bibinfo{year}{1997}), \bibinfo{pages}{56--58}.
\newblock


\bibitem[\protect\citeauthoryear{Rumelhart, Hinton, and Williams}{Rumelhart
  et~al\mbox{.}}{1985}]%
        {rumelhart1985learning}
\bibfield{author}{\bibinfo{person}{David~E Rumelhart},
  \bibinfo{person}{Geoffrey~E Hinton}, {and} \bibinfo{person}{Ronald~J
  Williams}.} \bibinfo{year}{1985}\natexlab{}.
\newblock \bibinfo{booktitle}{{\em Learning internal representations by error
  propagation}}.
\newblock \bibinfo{type}{{T}echnical {R}eport}.
  \bibinfo{institution}{California Univ San Diego La Jolla Inst for Cognitive
  Science}.
\newblock


\bibitem[\protect\citeauthoryear{Schafer, Konstan, and Riedl}{Schafer
  et~al\mbox{.}}{1999}]%
        {schafer1999recommender}
\bibfield{author}{\bibinfo{person}{J~Ben Schafer}, \bibinfo{person}{Joseph
  Konstan}, {and} \bibinfo{person}{John Riedl}.}
  \bibinfo{year}{1999}\natexlab{}.
\newblock \showarticletitle{Recommender systems in e-commerce}. In
  \bibinfo{booktitle}{{\em Proceedings of the 1st ACM conference on Electronic
  commerce}}. \bibinfo{pages}{158--166}.
\newblock


\bibitem[\protect\citeauthoryear{Schmidhuber}{Schmidhuber}{2015}]%
        {schmidhuber2015deep}
\bibfield{author}{\bibinfo{person}{J{\"u}rgen Schmidhuber}.}
  \bibinfo{year}{2015}\natexlab{}.
\newblock \showarticletitle{Deep learning in neural networks: An overview}.
\newblock \bibinfo{journal}{{\em Neural networks\/}}  \bibinfo{volume}{61}
  (\bibinfo{year}{2015}), \bibinfo{pages}{85--117}.
\newblock


\bibitem[\protect\citeauthoryear{Seide, Li, Chen, and Yu}{Seide
  et~al\mbox{.}}{2011}]%
        {seide2011feature}
\bibfield{author}{\bibinfo{person}{Frank Seide}, \bibinfo{person}{Gang Li},
  \bibinfo{person}{Xie Chen}, {and} \bibinfo{person}{Dong Yu}.}
  \bibinfo{year}{2011}\natexlab{}.
\newblock \showarticletitle{Feature engineering in context-dependent deep
  neural networks for conversational speech transcription}. In
  \bibinfo{booktitle}{{\em 2011 IEEE Workshop on Automatic Speech Recognition
  \& Understanding}}. IEEE, \bibinfo{pages}{24--29}.
\newblock


\bibitem[\protect\citeauthoryear{Shan, Hoens, Jiao, Wang, Yu, and Mao}{Shan
  et~al\mbox{.}}{2016}]%
        {shan2016deep}
\bibfield{author}{\bibinfo{person}{Ying Shan}, \bibinfo{person}{T~Ryan Hoens},
  \bibinfo{person}{Jian Jiao}, \bibinfo{person}{Haijing Wang},
  \bibinfo{person}{Dong Yu}, {and} \bibinfo{person}{JC Mao}.}
  \bibinfo{year}{2016}\natexlab{}.
\newblock \showarticletitle{Deep Crossing: Web-Scale Modeling without Manually
  Crafted Combinatorial Features}. In \bibinfo{booktitle}{{\em Proceedings of
  the 22nd ACM SIGKDD International Conference on Knowledge Discovery and Data
  Mining}}. ACM, \bibinfo{pages}{255--262}.
\newblock


\bibitem[\protect\citeauthoryear{Shazeer, Mirhoseini, Maziarz, Davis, Le,
  Hinton, and Dean}{Shazeer et~al\mbox{.}}{2017}]%
        {shazeer2017outrageously}
\bibfield{author}{\bibinfo{person}{Noam Shazeer}, \bibinfo{person}{Azalia
  Mirhoseini}, \bibinfo{person}{Krzysztof Maziarz}, \bibinfo{person}{Andy
  Davis}, \bibinfo{person}{Quoc Le}, \bibinfo{person}{Geoffrey Hinton}, {and}
  \bibinfo{person}{Jeff Dean}.} \bibinfo{year}{2017}\natexlab{}.
\newblock \showarticletitle{Outrageously large neural networks: The
  sparsely-gated mixture-of-experts layer}.
\newblock \bibinfo{journal}{{\em arXiv preprint arXiv:1701.06538\/}}
  (\bibinfo{year}{2017}).
\newblock


\bibitem[\protect\citeauthoryear{Song, Shi, Xiao, Duan, Xu, Zhang, and
  Tang}{Song et~al\mbox{.}}{2019}]%
        {song2019autoint}
\bibfield{author}{\bibinfo{person}{Weiping Song}, \bibinfo{person}{Chence Shi},
  \bibinfo{person}{Zhiping Xiao}, \bibinfo{person}{Zhijian Duan},
  \bibinfo{person}{Yewen Xu}, \bibinfo{person}{Ming Zhang}, {and}
  \bibinfo{person}{Jian Tang}.} \bibinfo{year}{2019}\natexlab{}.
\newblock \showarticletitle{Autoint: Automatic feature interaction learning via
  self-attentive neural networks}. In \bibinfo{booktitle}{{\em Proceedings of
  the 28th ACM International Conference on Information and Knowledge
  Management}}. \bibinfo{pages}{1161--1170}.
\newblock


\bibitem[\protect\citeauthoryear{Valiant}{Valiant}{2014}]%
        {valiant2014learning}
\bibfield{author}{\bibinfo{person}{Gregory Valiant}.}
  \bibinfo{year}{2014}\natexlab{}.
\newblock \showarticletitle{Learning polynomials with neural networks}.
\newblock  (\bibinfo{year}{2014}).
\newblock


\bibitem[\protect\citeauthoryear{Vaswani, Shazeer, Parmar, Uszkoreit, Jones,
  Gomez, Kaiser, and Polosukhin}{Vaswani et~al\mbox{.}}{2017}]%
        {vaswani2017attention}
\bibfield{author}{\bibinfo{person}{Ashish Vaswani}, \bibinfo{person}{Noam
  Shazeer}, \bibinfo{person}{Niki Parmar}, \bibinfo{person}{Jakob Uszkoreit},
  \bibinfo{person}{Llion Jones}, \bibinfo{person}{Aidan~N Gomez},
  \bibinfo{person}{{\L}ukasz Kaiser}, {and} \bibinfo{person}{Illia
  Polosukhin}.} \bibinfo{year}{2017}\natexlab{}.
\newblock \showarticletitle{Attention is all you need}. In
  \bibinfo{booktitle}{{\em Advances in neural information processing systems}}.
  \bibinfo{pages}{5998--6008}.
\newblock


\bibitem[\protect\citeauthoryear{Veit, Wilber, and Belongie}{Veit
  et~al\mbox{.}}{2016}]%
        {NIPS2016_6556}
\bibfield{author}{\bibinfo{person}{Andreas Veit}, \bibinfo{person}{Michael~J
  Wilber}, {and} \bibinfo{person}{Serge Belongie}.}
  \bibinfo{year}{2016}\natexlab{}.
\newblock \showarticletitle{Residual Networks Behave Like Ensembles of
  Relatively Shallow Networks}.
\newblock In \bibinfo{booktitle}{{\em Advances in Neural Information Processing
  Systems 29}}, \bibfield{editor}{\bibinfo{person}{D.~D. Lee},
  \bibinfo{person}{M.~Sugiyama}, \bibinfo{person}{U.~V. Luxburg},
  \bibinfo{person}{I.~Guyon}, {and} \bibinfo{person}{R.~Garnett}} (Eds.).
  \bibinfo{publisher}{Curran Associates, Inc.}, \bibinfo{pages}{550--558}.
\newblock


\bibitem[\protect\citeauthoryear{Wang, Fu, Fu, and Wang}{Wang
  et~al\mbox{.}}{2017}]%
        {wang2017deep}
\bibfield{author}{\bibinfo{person}{Ruoxi Wang}, \bibinfo{person}{Bin Fu},
  \bibinfo{person}{Gang Fu}, {and} \bibinfo{person}{Mingliang Wang}.}
  \bibinfo{year}{2017}\natexlab{}.
\newblock \showarticletitle{Deep \& Cross Network for Ad Click Predictions}.
\newblock In \bibinfo{booktitle}{{\em Proceedings of the ADKDD'17}}.
  \bibinfo{pages}{1--7}.
\newblock


\bibitem[\protect\citeauthoryear{Wang, Li, Mahoney, and Darve}{Wang
  et~al\mbox{.}}{2019}]%
        {wang2019block}
\bibfield{author}{\bibinfo{person}{Ruoxi Wang}, \bibinfo{person}{Yingzhou Li},
  \bibinfo{person}{Michael~W Mahoney}, {and} \bibinfo{person}{Eric Darve}.}
  \bibinfo{year}{2019}\natexlab{}.
\newblock \showarticletitle{Block Basis Factorization for Scalable Kernel
  Evaluation}.
\newblock \bibinfo{journal}{{\it SIAM J. Matrix Anal. Appl.}}
  \bibinfo{volume}{40}, \bibinfo{number}{4} (\bibinfo{year}{2019}),
  \bibinfo{pages}{1497--1526}.
\newblock


\bibitem[\protect\citeauthoryear{Yu, Liu, Wang, and Tao}{Yu
  et~al\mbox{.}}{2017}]%
        {yu2017compressing}
\bibfield{author}{\bibinfo{person}{Xiyu Yu}, \bibinfo{person}{Tongliang Liu},
  \bibinfo{person}{Xinchao Wang}, {and} \bibinfo{person}{Dacheng Tao}.}
  \bibinfo{year}{2017}\natexlab{}.
\newblock \showarticletitle{On compressing deep models by low rank and sparse
  decomposition}. In \bibinfo{booktitle}{{\em Proceedings of the IEEE
  Conference on Computer Vision and Pattern Recognition}}.
  \bibinfo{pages}{7370--7379}.
\newblock


\end{thebibliography}

\clearpage
{\LARGE \bf Appendix}
\setlength{\belowdisplayskip}{0pt} \setlength{\belowdisplayshortskip}{0pt}
\setlength{\abovedisplayskip}{0pt} \setlength{\abovedisplayshortskip}{0pt}
\allowdisplaybreaks

\section{Baseline performance reported in papers}
\label{sec:metrics_reported}
Tab. \ref{tab:metrics_in_papers} lists the quoted Logloss and AUC metrics reported in papers for each baseline. 
\begin{table*}[htbp]
\small
\caption{Baseline performance reported in papers. The metrics (Logloss, AUC) are quoted from papers. Each row represents a baseline, each column represents the paper where the metrics are being reported. The best metric for each baseline is marked in bold.}
\vspace{-3.5ex}
\label{tab:metrics_in_papers}
\begin{center}
\begin{tabular}{c|cccccc}
\toprule
\diagbox[width=\dimexpr \textwidth/8+2\tabcolsep\relax, height=0.5cm]{ \bf Model }{\bf Paper} & DeepFM\cite{guo2017deepfm} (2017)& DCN\cite{wang2017deep} (2017) & xDeepFM\cite{lian2018xdeepfm} (2018) & DLRM\cite{naumov2019deep} (2019) & AutoInt\cite{song2019autoint} (2019) & {DCN-V2} (ours)\\
\midrule
DeepFM  & (0.45083, 0.8007) &  -- & (0.4468, 0.8025) & -- & (0.4449, 0.8066) & \bf (0.4420, 0.8099)\\
DCN     & -- & (0.4419, -) & (0.4467, 0.8026) & (-, $\sim$ 0.789) & (0.4447, 0.8067) & \bf (0.4420, 0.8099) \\
xDeepFM & -- & -- & ({\bf 0.4418}, 0.8052) & -- & (0.4447, 0.8070) & (0.4421, \bf{0.8099})\\
DLRM & -- & -- & -- & (-, $\sim$ 0.790) & -- & (0.4427, \bf{0.8092}) \\
AutoInt & -- & -- & -- & -- & (0.4434, 0.8083) & \bf (0.4420, 0.8101)\\
DCN-V2 & -- & -- & -- & -- & -- & \bf (0.4406, 0.8115)\\
DNN & -- & (0.4428, -) & (0.4491, 0.7993) & -- & -- & \bf (0.4421, 0.8098)\\
\bottomrule
\end{tabular}
\end{center}
\end{table*}

\section{Theorem Proofs}
\subsection{Proofs for \autoref{thm:cross_x0_featurewise}}
\begin{proof}

\begin{itemize}[leftmargin=0em]
We start with notations; then prove by induction.
\item[] {\bf Notations.}
Let $[k] := \{1, \ldots, k\}$. Let's denote the embedding as $\vecx = [\vecx_1; \vecx_2; \ldots; \vecx_c]$, the output from the $l$-th cross layer to be $\vecx^l = [\vecx_1^l; \vecx_2^l; \ldots; \vecx_c^l]$ where $\vecx_i, \vecx_i^l \in \mathbb{R}^{e_i}$ and $e_i$ is the embedding size for the $i$-th feature. 
To simplify the notations, let's also define the feature interaction between features in an ordered set $I$ (\emph{e.g.,} $(i_1, i_3, i_4)$) with weights characterized by an ordered set $J$ as
\begin{equation}
\label{eq:interaction_g}
g(I, J; \vecx, W) = \vecx_{i_1} \odot \left(W_{i_1, i_2}^{j_1} \vecx_{i_2} \odot \ldots \odot \left(W_{i_k, i_{k+1}}^{j_{k}} \vecx_{i_{l+1}}\right) \right)
\end{equation}
where weights $W_{i_a, i_b}^j$ represents the $(i_a, i_b)$-th block in weight $W^j$ at the $j$-th cross layer, and it serves as two purposes: align the dimensions between features and increase the impressiveness of the feature cross representations. Note that given the order of $\vecx_i$'s, the subscripts of matrix $W$'s are uniquely determined.

\item[] {\bf Proposition.} We first proof by induction that $\vecx_i^l$ has the following formula:
\begin{equation}
\label{eq:general_form_x_i}
\vecx_i^l = \sum_{p=2}^{l+1} \sum_{I \in S_p^i} \sum_{J \in C_l^{p-1}} g(I, J; \vecx, W) + \vecx_{i}
\end{equation}
where $S_p^i$ is a set which represents all the combinations of choosing $p$ elements from $[c]$ with replacement, and with first element fixed to be $i$: $S_p^i =: \bigl\{\vecy \in [c]^{p} \mathrel{\big|} y_1 = i\bigr\}, ~\forall I \in S_p, ~I = (i_1, \ldots, i_p);$
and $C_l^{p-1}$ is a set that represents choosing a combination of $p-1$ indices out of integers $[l]$ at a time:
$C_l^{p-1} := \bigl\{\vecy \in [l]^{p-1} \mathrel{\big|} \forall i < j, y_i > y_j  \bigr\}.$

\item[] {\bf Base case.} When $l=1$, $\vecx_i^1 = \sum_j W_{i,j}^{1} \vecx_j + \vecx_i$.

\item[] {\bf Induction step.} Let's assume that when $l = k$, 
$$\vecx_i^k = \sum_{p=2}^{k+1} \sum_{I \in S_p^i} \sum_{J \in C_k^{p-1}} g_J(\vecx; I) + \vecx_i
$$
Then, for $l = k+1$, we have 
\begin{align*}
&\vecx_{i}^{k+1} = \vecx_i \odot \sum_{q=1}^c W_{i, q}^{k+1} \vecx_q^k + \vecx_i^k \\ 
=& ~\vecx_i \odot \sum_{q=1}^c W_{i, q}^{k+1} \left(\sum_{p=2}^{k+1} \sum_{I \in S_p^q} \sum_{J \in C_k^{p-1}} g(I, J; \vecx, W) + \vecx_q\right) + \\
& \sum_{p=2}^{k+1} \sum_{I \in S_p^i} \sum_{J \in C_k^{p-1}} g(I, J; \vecx, W) + \vecx_i \\
=& \sum_{q=1}^c \sum_{p=2}^{k+1} \sum_{I \in S_p^q} \sum_{J \in C_k^{p-1}} \vecx_i \odot \left(W_{i, q}^{k+1} g(I, J; \vecx, W)\right) + \\ 
& \sum_{q=1}^c \vecx_i \odot W_{i, q}^{k+1}\vecx_q+  \sum_{p=2}^{k+1} \sum_{I \in S_p^i } \sum_{J \in C_k^{p-1}} g(I, J; \vecx, W) + \vecx_i\\ 
=&  \sum_{p=2}^{k+1} \sum_{J \in C_k^{p-1}} \sum_{q=1}^c \sum_{I \in S_p^q} \vecx_i \odot \left(W_{i, q}^{k+1} g(I, J; \vecx, W)\right) +\\
& \sum_{p=2} \sum_{J=k+1} \sum_{I \in S_2^i}g(I, J; \vecx, W)+ \sum_{p=2}^{k+1} \sum_{I \in S_p^i } \sum_{J \in C_k^{p-1}} g(I, J; \vecx, W) + \vecx_i\\ 
=&  \sum_{p=2}^{k+1} \sum_{J \in {k+1} \oplus C_k^{p-1}} \sum_{I \in S_{p+1}^i} g(I, J; \vecx, W)+\\
& \sum_{p=2} \sum_{J=k+1} \sum_{I \in S_2^i}g(I, J; \vecx, W)+ \sum_{p=2}^{k+1} \sum_{I \in S_p^i } \sum_{J \in C_k^{p-1}} g(I, J; \vecx, W) + \vecx_i \\ 
=&  \left(\sum_{p=3}^{k+2} \sum_{J \in {k+1} \oplus C_k^{p-2}} \sum_{I \in S_{p}^i}+\sum_{p=3}^{k+1} \sum_{I \in S_p^i } \sum_{J \in C_k^{p-1}}\right) g(I, J; \vecx, W)+\\
& \left(\sum_{p=2} \sum_{I \in S_2^i } \sum_{J \in C_k^1} g(I, J; \vecx, W) + \sum_{p=2} \sum_{J=k+1} \sum_{I \in S_2^i} \right)g(I, J; \vecx, W)+\vecx_i\\
=& \sum_{p=3}^{k+2} \sum_{J \in C_{k+1}^{p-1}} \sum_{I \in S_{p}^i}g(I, J; \vecx, W) +\sum_{p=2} \sum_{J=C_{k+1}^{p-1}} \sum_{I \in S_p^i} g(I, J; \vecx, W)+\vecx_i\\ 
=& \sum_{p=2}^{k+2} \sum_{I \in S_p^i } \sum_{J \in C_{k+1}^{p-1}} g(I, J; \vecx, W) + \vecx_i
\end{align*}
where $\oplus$ denotes adding index $k+1$ to each element in the set of $C_{k}^{p-1}$. The first $5$ equalities are are straightforward.
For the $6^\text{th}$ equality, we first interchanged variable $p' = p+1$ for the first term, and separated the third term into cases of $p=2$ and $p > 2$. Then, we group the terms into two cases: $p=2$ and $p>2$. 
For the second to the last equality, we combined the summations over $J$. Consider the set of choosing a combination of $p-1$ indices from $k+1$ integers, it could be separated into two sets, with index $k+1$ and without. Hence, $C_{k+1}^{p-1} = C_{k}^{p-1} \cup \left((k+1) \oplus C_{k}^{p-2}\right)$. 

\item[] {\bf Conclusion.} Since both the base case and the induction step hold, we conclude that $\forall~ l \ge 1$, Eq \eqref{eq:general_form_x_i} holds. This completes the proof.

In such case, the $l$-th cross layer contains all the feature interactions (feature-wise) of order up to $l+1$. The interactions between different feature set is parameterized differently, specifically, the interactions between features in set $I$ (feature's can be repeated) of order $p$ is 
\begin{equation*}
\begin{split}
\sum_{{\bf i} \in I'} \sum_{\vecj \in C_p^{p-1}} \left\{g(\veci, \vecj; \vecx, W)
= \vecx_{i_1} \odot \left(W_{i_1, i_2}^{j_1} \vecx_{i_2} \odot \ldots \odot \left(W_{i_k, i_{k+1}}^{j_{k}} \vecx_{i_{l+1}}\right) \right)\right\}
\end{split}
\end{equation*}
where $I'$ contains all the permutations of elements in $I$.

\end{itemize}
\end{proof}

\subsection{Proofs for \autoref{thm:cross_x0_bitwise}}
\begin{proof}
Instead of treating each feature embedding as a unit, we treat each element $x_i$ in input embedding $\vecx = [x_1, x_2, \ldots, x_d]$ as a unit. This is a special case of \autoref{thm:cross_x0_featurewise} where all the feature embedding sizes are 1.  In such case, all the computations are interchangeable. Hence, we adopt the notations and also the result of \autoref{eq:general_form_x_i}, that is, the $i$-th element in the $l$-th layer of cross network $\vecx^l$ has the following formula:
\begin{equation}
\label{eq:general_form_x_i_bit}
\vecx_i^l = \sum_{p=2}^{l+1} \sum_{I \in S_p^i} \sum_{J \in C_l^{p-1}} g(I, J; \vecx, W) + x_i
\end{equation}

To ease the proof and simplify the final formula, we assume the final logit for a $l$-layer cross network is ${\bf 1}^\top \vecx^l$, then
\begin{equation*}
\begin{split}
    {\bf 1}^\top \vecx^l &= \sum_{i=1}^d \sum_{p=2}^{l+1} \sum_{I \in S_p^i} \sum_{J \in C_l^{p-1}} x_{i_1} \odot \left(w_{i_1i_2}^{(j_1)} x_{i_2} \odot \ldots \odot \left(w_{i_ki_{k+1}}^{(j_{k})} x_{i_{l+1}}\right) \right) + \sum_{i=1}^dx_i\\
    &= \sum_{p=2}^{l+1} \sum_{I \in S_p} \sum_{J \in C_l^{p-1}}  w_{i_1i_2}^{(j_1)} \ldots w_{i_ki_{k+1}}^{(j_{k})} x_{i_1} x_{i_2}  \ldots x_{i_{l+1}} +  \sum_{i=1}^dx_i\\
    &=  \sum_{p=2}^{l+1} \sum_{|\vecalpha| = p} \sum_{J \in C_l^{p-1}}  \sum_{\veci \in P_\vecalpha} \prod_{k=1}^{|\vecalpha|-1} w_{i_k i_{k+1}}^{(j_k)} x_1^{\alpha_1}x_2^{\alpha_2}\cdots x_d^{\alpha_d} +  \sum_{i=1}^dx_i\\
    &=   \sum_{\vecalpha} \sum_{\vecj \in C_l^{|\vecalpha|-1}}  \sum_{\veci \in P_\vecalpha} \prod_{k=1}^{|\vecalpha|-1} w_{i_k i_{k+1}}^{(j_k)} x_1^{\alpha_1}x_2^{\alpha_2}\cdots x_d^{\alpha_d}+  \sum_{i=1}^dx_i
\end{split}
\end{equation*}
where $P_\vecalpha$ is the set of all the permutations of $(\underbrace{1 \cdots 1}_{\alpha_1~ \text{times}} \cdots \underbrace{d \cdots d}_{\alpha_d~ \text{times}})$, $C_l^{|\vecalpha|-1}$ is a set that represents choosing a combination of $|\vecalpha|-1$ indices out of integers $\{1,\cdots, l\}$ at a time, specifically,
$$C_l^{|\vecalpha|-1} \coloneqq \bigl\{\vecy \in [l]^{|\vecalpha|-1} \mathrel{\big|} (y_i \neq y_j) ~\wedge~ (y_{j_1} > y_{j_2} > \ldots > y_{j_{|\vecalpha|-1}}) \bigr\}.$$

The second equality combined the first and the third summations into a single one summing over a new set $S_p^c := [c]^{p}$. The third equality re-represented the cross terms (monomials) using multi-index $\vecalpha$, and modified the index for weights $w$'s accordingly. The last equality combined the first two summations. Thus the proof.
\end{proof}

\end{document}